\documentclass[]{natureprintstyle}
\usepackage{graphicx}
\usepackage{longtable}
\usepackage{amssymb}

\usepackage{multirow}
\usepackage{savesym}
\usepackage{amsmath}
\savesymbol{iint}
\usepackage{txfonts}
\restoresymbol{TXF}{iint}
\usepackage{graphicx}
\usepackage{ulem} 
\usepackage{url}
\usepackage{threeparttable}
\usepackage{blindtext}
\usepackage{xcolor}
\usepackage{soul}
\usepackage{booktabs}
\usepackage[super,sort,comma]{natbib}
\usepackage{bibunits}

\usepackage{etoolbox}
\makeatletter
\newcommand*{\newbibstartnumber}[1]{%
  \apptocmd{\thebibliography}{%
    \global\c@NAT@ctr #1\relax
    \addtocounter{NAT@ctr}{-1}%
  }{}{}%
}
\makeatother

\usepackage[colorlinks, linkcolor=violet, anchorcolor=green, citecolor=blue]{}

\def\arcsec{\hbox{$^{\prime\prime}$}}
\def\urlprefix   {{\sc url: }}

\def\purple#1 {{\textcolor{purple}{#1}}\ }
\def\red#1 {\textcolor{red}{#1}}
\def\new#1 {{\bf #1 }}
\def\emph#1  {\textit{ #1 } }%

\newcommand{\apj}{Astrophys. J.}
\newcommand{\procspie}{Proc. SPIE}
\newcommand{\pasp}{Publ. Astron. Soc. Pac.}
\newcommand{\apjs}{Astrophys. J. Supp.}
\newcommand{\araa}{Annu. Rev. Astron. Astrophys.}
\newcommand{\mnras}{Mon. Not. R. Astron. Soc.}
\newcommand{\apjl}{Astrophys. J. Let.}
\newcommand{\aap}{Astron. Astrophys.}

\newcommand{\nat}{Nature}

\newcommand{\pasj}{Publ. Astron. Soc. Jpn.}

\newcommand{\gt}{$>$}
\newcommand{\lt}{$<$}
\newcommand{\amp}{$\&$}

\title{\bf A dominant population of optically invisible massive galaxies in the early Universe}

\author{T.~Wang$^{1,2,3}$, C.~Schreiber$^{4,5,6}$, D.~Elbaz$^{2}$, Y.~Yoshimura$^{1}$, K.~Kohno$^{1,6}$, X.~Shu$^{7}$, Y.~Yamaguchi$^{1}$, M.~Pannella$^{8}$, M.~Franco$^{2}$,  J.~Huang$^{9}$, C.-F. Lim$^{10,11}$ \& W.-H.~Wang$^{10}$}

\begin{document}

\begin{bibunit}[plainnat]

\maketitle

\begin{affiliations}
\item Institute of Astronomy, Graduate School of Science,
The University of Tokyo, Tokyo, Japan
\item AIM, CEA, CNRS, Universit\'e Paris Diderot, Saclay, Sorbonne Paris Cit\'e, Gif-sur-Yvette, France
\item National Astronomical Observatory of Japan, Mitaka, Tokyo, Japan
\item Leiden Observatory, Leiden University, Leiden, The Netherlands
\item Department of Physics, University of Oxford, Oxford, UK
\item Research Center for the Early Universe, Graduate
School of Science, Tokyo, Japan
\item Department of Physics, Anhui Normal University, Wuhu, China
\item Faculty of Physics, Ludwig-Maximilians-Universit{\"a}t, Munich, Germany
\item National Astronomical Observatories of China, Chinese Academy of Sciences, Beijing, China
\item Academia Sinica Institute of Astronomy and Astrophysics, Taipei,, Taiwan
\item Graduate Institute of Astrophysics, National Taiwan University, Taipei, Taiwan

\end{affiliations}
\begin{abstract}

Our current knowledge of cosmic star-formation history during the first two billion years (corresponding to redshift $z >3$) is mainly based on galaxies identified in rest-frame ultraviolet light\cite{Madau:2014}. However, this population of galaxies is known to under-represent the most massive galaxies, which have rich dust content and/or old stellar populations. This raises the questions of the true abundance of massive galaxies and the star-formation-rate density in the early universe. Although several massive galaxies that are invisible in the ultraviolet have recently been confirmed at early epochs\cite{Walter:2012,Riechers:2013,Marrone:2018}, most of them are extreme starbursts with star-formation rates exceeding 1000 solar masses per year, suggesting that they are unlikely to represent the bulk population of massive galaxies.
Here we report submillimeter (wavelength 870~$\mu$m) detections of 39 massive star-forming galaxies at $z > 3$, which are unseen in the spectral region from the deepest ultraviolet to the near-infrared. With a space density of about $2 \times 10^{-5}$ per cubic megaparsec (two orders of magnitudes higher than extreme starbursts\cite{Dowell:2014}) and star-formation rates of $\sim$200 solar masses per year, these galaxies represent the bulk population of massive galaxies that have been missed from previous surveys.  They contribute a total star-formation- rate density ten times larger than that of equivalently massive ultraviolet-bright galaxies at $z >3$. 
Residing in the most massive dark matter halos at their redshifts, they are probably the progenitors of the largest present-day galaxies in massive groups and clusters.
Such a high abundance of massive and dusty galaxies in the early universe challenges our understanding of massive-galaxy formation. 
\end{abstract}

Observations of galaxies across cosmic time have revealed that more massive galaxies have assembled their stellar masses at earlier epochs, with a significant population of massive ellipticals already in place at redshifts $z \sim 3-4$~\cite{Glazebrook:2017,Schreiber:2018c,Spitler:2014}. The early assembly of these massive galaxies has posed serious challenges to current galaxy formation theories. Understanding their formation processes  requires studies of their progenitors formed at even higher redshifts. However, most currently known high-redshift galaxies, including mainly Lyman-break galaxies (LBGs) and few extreme starbursts, are found inadequate to account for the large population of these early formed ellipticals, due to either low stellar masses and star formation rates, SFRs (for LBGs\cite{Williams:2014}) or low space densities (for the extreme starbursts). This suggests that the main progenitors of massive galaxies at $z > 3$ remain to be found. Identification of these currently missing massive galaxies is key to our understanding of both massive-galaxy formation and  the cosmic SFR density in the early universe.

\begin{figure*}
\centering
\includegraphics[scale=0.7]{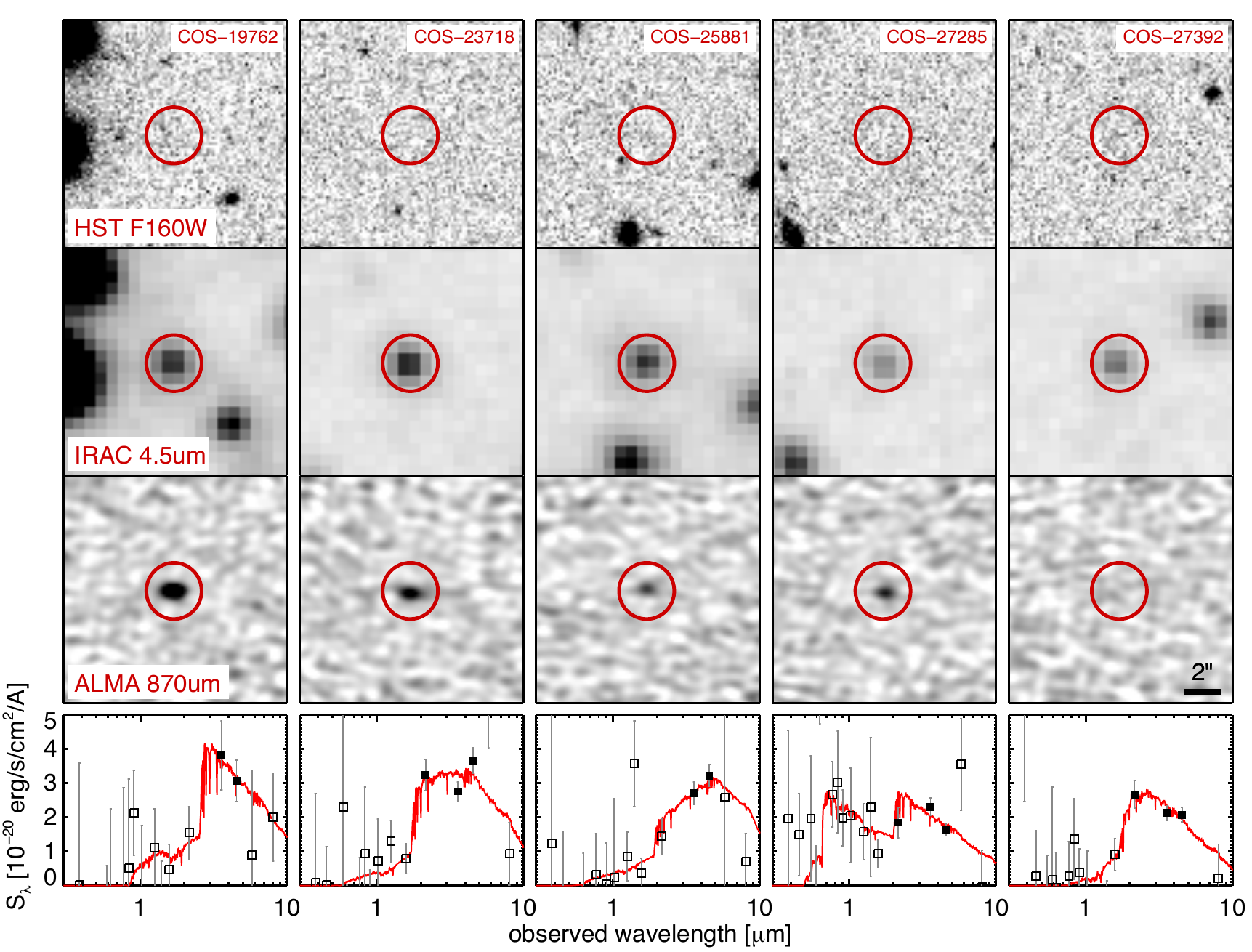}
\caption{\textbf{Example images and UV-to-NIR SEDs of H-dropouts.} Top three rows, images of five H-dropouts obtained in three different spectral bands-HST/F160W (top row), IRAC 4.5 $\mu$m (second row), and ALMA 870 $\mu$m (third row). The H-dropouts, named in the top row, were selected randomly from the parent sample, with all but the last one (COS-27392) detected with ALMA. Each image is 12$\arcsec \times 12\arcsec$; see scale bar in bottom right image. Bottom row, the measured UV-to-NIR SED (squares) and best-fit stellar population synthesis models (red lines). The error bars are 1 $\sigma$. The filled and open squares indicate photometric points with measured signal-to-noise ratio (S/N) above and below 3, respectively. 
\label{Fig:stamps}
}
\end{figure*}

The main targets of this study are a population of galaxies that are {\it Spitzer}/Infrared Array Camera (IRAC)-bright yet undetected in even the deepest near-infrared (NIR: H-band) imaging with Hubble Space Telescope ({\it HST}), that is, H-dropouts. (Throughout this Letter we use the short form ``Telescope/Instrument'' to represent usage of a particular instrument on a particular telescope.) In total, we have identified 63 $H-$dropouts with IRAC 4.5-$\mu$m magnitude, [4.5], less than 24 mag, within a total survey area of $\sim 600$ arcmin$^{2}$ in deep CANDELS fields with typical depth of {\it H} $>27$ mag (5$\sigma$)(Fig.~\ref{Fig:stamps}, Extended Data Table 1, Methods). 
Although previous studies have shown that these bright and red IRAC sources are promising candidates for massive galaxies at\cite{HuangJ:2011,WangT:2016a} $z > 3$, confirming their nature has been difficult so far owing to the limited sample size, the poor resolution of {\it Spitzer} and the lack of multiwavelength information. 
Here we explore their nature with high-resolution, 870 $\mu$m continuum imaging with the Atacama Large Millimeter/submillimeter Array (ALMA). With only 1.8 min of integration per object, 39 of them (detection rates of 62\%) are detected down to an integrated flux of 0.6 mJy (4$\sigma$, Extended Data Fig. 1,  Extended Data Table 2).
Their 870-$\mu$m fluxes range from 0.6 mJy to 8 mJy, with a median of $S_{870~\mu m}$ =1.6 mJy (Extended Data Fig. 2). Hence most of them are fainter than the 2-mJy confusion limit of the single dish instruments that discovered submillimeter galaxies (SMGs), and much fainter than most SMGs studied until now with typical\cite{Swinbank:2014} $S_{870~\mu m} \gtrsim 4$ mJy. The sky density of these ALMA-detected H-dropouts is  approximately $5.3 \times 10^{2}$ deg$^{-2}$ after correction for incompleteness (Methods), two orders of magnitude higher than \textit{Herschel}/SPIRE-selected extreme starbursts (with SFR $\gtrsim 1000~M_{\odot}$ yr$^{-1}$)~\cite{Riechers:2013,Dowell:2014}.

\begin{figure}
\centering
\includegraphics[scale=0.8]{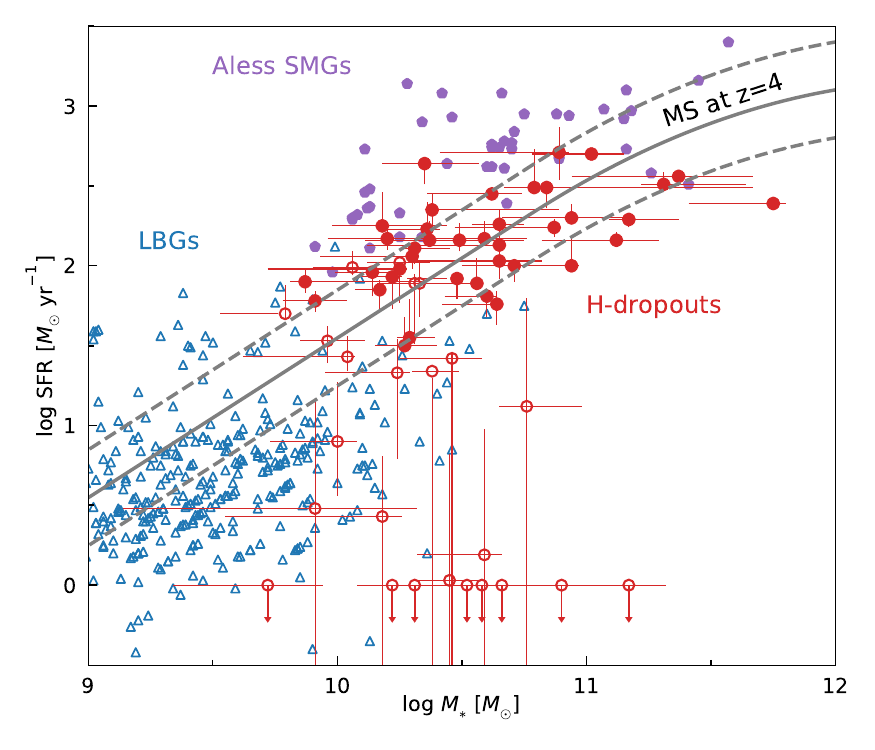}

\caption{\textbf{Stellar masses and star formation rates of H-dropouts.} The red filled and open circles represent respectively the ALMA-detected and ALMA-undetected $H-$dropouts. For comparison, a sample of LBGs at $z=4-6$ from the  ZFOURGE survey~\cite{Straatman:2016} and bright $z > 3$ SMGs ($S_{870 \mu m} > 4.2$ mJy) from the ALESS survey are also shown\cite{daCunha:2015}. The stellar masses for the ALESS SMGs are reduced by 0.3 dex to account for the systematic differences caused by the different methods used in mass estimation. The grey solid and dashed lines indicate respectively the star-forming main sequence (MS) at $z = 4$ and its 1$\sigma$ scatter\cite{Schreiber:2015}. 
The SFRs for ALMA-detected H-dropouts  are derived from the 870-$\mu$m fluxes assuming their intrinsic far-infrared SED resembles that of the stacked one. Error bars are 1$\sigma$. 
The SFRs for ALMA-undetected  H-dropouts are derived from UV-to-NIR SED fitting with an additional constraint of  SFR $>$ 1 $M_{\odot}$ yr$^{-1}$, for which  error bars represent the 16th and 84th percentiles of the distribution obtained in the Monte Carlo simulations (Methods), the same as that for stellar mass estimates.
\label{Fig:MS}
}
\end{figure}

The ALMA detections confirm unambiguously that most of the H-dropouts are dusty star-forming galaxies at high redshifts, consistent with their admittedly uncertain photometric redshifts--from optical spectral energy distribution (SED) fitting--with median redshift $z_{\rm median}=4$ (Extended Data Fig. 3). Further insights into their properties are obtained from the stacked infrared (IR) SED of the 39 ALMA-detected H-dropouts from MIPS 24 $\mu$m up to ALMA 870~$\mu$m. The stacked SED peaks between the observed 350 and 500 $\mu$m (Extended Data Figure 3), consistent with being at $z \sim 4$.  With a median stellar mass  of $M_{*} \sim 10^{10.6} M_{\odot}$ and a characteristic IR luminosity (over $8 - 1000$ $\mu$m) of $L_{\it IR} = 2.2\pm0.3 \times 10^{12} L_{\odot}$ ($L_{\odot}$, solar luminosity) derived from the stacked SED, these ALMA-detected H-dropouts are fully consistent with being normal massive star-forming galaxies at\cite{Schreiber:2018a}  $z=4$ (Fig.~\ref{Fig:MS}). Moreover, the ALMA detections also provide crucial constrains on the redshift of individual galaxies. Combined with SCUBA-2 450 $\mu$m and VLA 3 GHz data, the majority of the ALMA-detected H-dropouts exhibit  red $S_{870 \mu m}/S_{450 \mu m}$ and $S_{1.4 GHz}/S_{870 \mu m}$ colors that are suggestive of redshifts of $z > 3$ (Extended Data Fig. 4). Similarly, the non-detections at 24 $\mu$m (5$\sigma$ detection limit of 20 $\mu$Jy) for most of the sources implies red $S_{870~\mu m}/S_{24 \mu m}$ colors that are also consistent with $z > 3$ assuming typical SED templates\cite{Daddi:2009a}. 
We hence conclude that whereas the estimated redshifts for individual galaxies exhibit a large uncertainty, all the available data points to the ALMA-detected H-dropouts being massive, dusty star-forming galaxies at $z > 3$.

For the remaining approximately $40\%$ of H-dropouts that are not detected with ALMA, photometric redshift estimates based on their optical SEDs suggest a similar redshift distribution to that of ALMA-detected ones, with $z_{\rm median}=3.8$(Extended Data Figure 2). Their stacked ALMA 870 $\mu$m image yields a $6\sigma$ detection with $S_{870 \mu m} = 0.24\pm0.04$ mJy, approximately 8 times lower than that of ALMA-detected ones, suggesting lower specific SFRs compared to ALMA-detected ones, which is also confirmed by a full fitting of the stacked optical-to-IR SEDs (Extended Data Fig. 5).

\begin{figure*}
\centering
\includegraphics[scale=1.4]{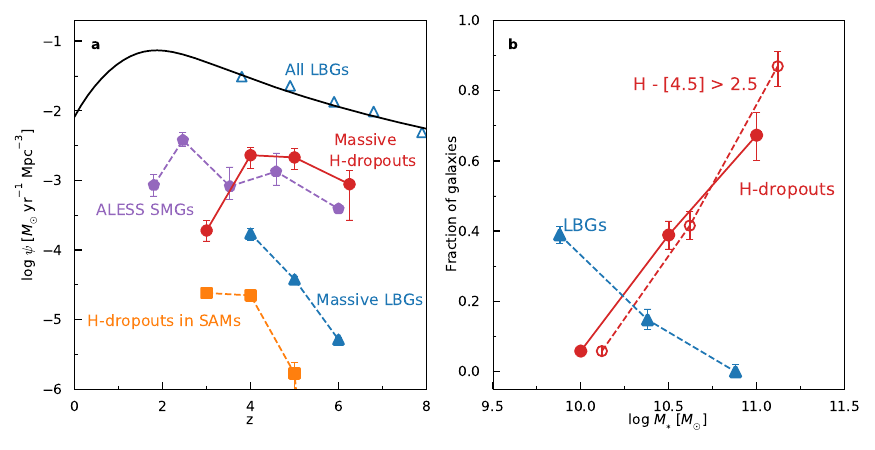}
\caption{\textbf{Contribution of H-dropouts to the cosmic SFR density and the stellar mass function.} {\bf a}: Plot of cosmic star-formation-rate density, $\psi$, versus redshift $z$. The black line indicates the current known total cosmic star-formation history, which is based on LBGs at $z \gtrsim 4$ ('All LBGs', blue open triangles~\cite{Bouwens:2012}). Red filed circles ('Massive H-dropouts'), ALMA-detected H-dropouts with $M_{*} > 10^{10.3} M_{\odot}$. Purple fileld pentagons, the ALESS SMGs ($S_{870 \mu m} > 4.2$ mJy)\cite{Swinbank:2014}, whose contribution to the SFR density peaks at $z \approx 2.5$. Blue filled triangles ('Massive LBGs'), the SFR density (based on dust-corrected UV) for the brightest/massive LBGs with $M_{*} > 10^{10.3} M_{\odot}$, based on the latest determination of the UV luminosity functions\cite{Ono:2018}. Filled orange squares, the SFR density from H-dropouts ($[4.5] < 24$ and $H - [4.5] > 2.5$) in semi-analytical models\cite{Henriques:2015}, which are identified from a K-selected mock catalog ($K < 27$) from a total area of 75.36 deg$^{2}$. Error bars, s.d. assuming Poisson statistics. {\bf b}: Number fraction of massive galaxies from the H-dropout sample and ZFOURGE catalogues that are detected either as LBGs (blue filled triangles) or H-dropouts (including both ALMA-detected and ALMA-undetected ones; red filled circles) averaged over $z=3.5-6.5$. Red open circles, the total contribution of red galaxies, including both H-dropouts and those non-H-dropouts that have similar red colors ($H-[4.5] > 2.5$) selected from ZFOURGE at $3.5<z<6.5$.
\label{Fig:SFRD}
}
\end{figure*}

Spectroscopic confirmation of H-dropouts has been so far limited to a few sources, which are all found at $z > 3$. Most of these confirmed cases are extreme SMGs with $S_{870 \mu m} \gtrsim 10$ mJy, for example$\cite{Walter:2012}., HDF-850 at $z = 5.18. An {\it H}-dropout galaxy with submillimeter flux similar to that of our sample ($S_{744 \mu m} = 2.3 \pm 0.1$ mJy) has been recently confirmed\cite{Schreiber:2018b} to be at $z = 3.709$: it was discovered serendipitously near a quiescent galaxy at the same redshift~\cite{Glazebrook:2017}. By targeting 3 H-dropouts in our sample that show significant excess ($> 4\sigma$, Methods) in Subaru medium bands in the optical ($\sim$ 3500-6000 \AA) with VLT/X-shooter, we have successfully detected Lyman-$\alpha$ for two of them and confirmed their redshifts to be $z > 3$ ($z = 3.097$ and $z=5.113$, Extended Data Fig. 6). These spectroscopic redshifts ($z_{\rm spec}$ are in good agreement with their photometric redshift ($z_{\rm phot}$) based on UV-to-NIR SED fitting, with $\sigma_{\Delta z/(1+z_{spec})} \sim 0.1$.

Having established that most of the H-dropouts are massive galaxies at $z > 3$, we now derive their contribution to the cosmic SFR density and stellar mass function. Whereas populations of similarly red galaxy populations are known to exist at lower redshifts\cite{Riguccini:2015}, these  largely overlap with the stellar-mass-limited sample used to estimate the SFR density at $z < 3$.
Assuming that the intrinsic infrared SED of the ALMA-detected H-dropouts is the same as the SED derived from stacking, 
the SFR density of ALMA-detected H-dropouts (in 10$^{-3} M_{\odot}$ yr$^{-1}$ Mpc$^{-3}$) reaches about 2.9, 2.1, and 0.9 at $z=$ 4, 5, 6, respectively, or approximately 1.6$\times 10^{-3} M_{\odot}$ yr$^{-1}$ Mpc$^{-3}$ when averaged over the three bins (Fig.~\ref{Fig:SFRD}). This corresponds to about 10\% of the SFR density from LBGs at similar redshifts\cite{Bouwens:2012}. However, if we focus only on LBGs with masses similar to those of $H-$dropouts with $M_{*} > 10^{10.3} M_{\odot}$, the SFR densities of $H-$dropouts are one to two orders of magnitude higher, demonstrating that H-dropouts dominate the SFR density in massive galaxies. This dominance is further reflected in the stellar mass functions, as shown in Fig.~\ref{Fig:SFRD}. The fraction of $H$-dropout becomes progressively higher at higher masses. At $M_{*} \gtrsim 10^{10.5} M_{\odot}$, the number density of $H$-dropout surpasses that of LBGs. Moreover, if we also include galaxies detected in $H-$band but which show similar red colors ($H - [4.5] > 2.5$, Extended Data Fig. 7)\cite{Spitler:2014, WangT:2016a}, they make up more than than 80\% of the most massive galaxies at $z > 4$. Taken together, these results suggest that the majority of the most massive galaxies at $z > 3$ have  indeed been missed from the LBG selection, and are optically dark.

\begin{figure}
\centering
\includegraphics[scale=0.95]{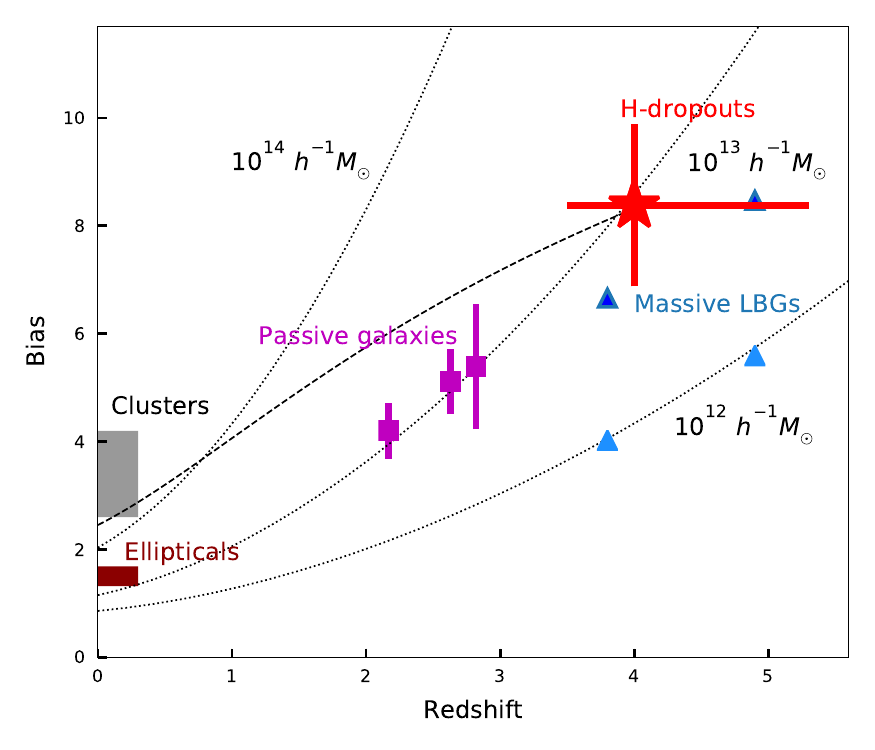}
\caption{\textbf{Clustering properties and halo masses of H-dropouts.} Shown is the galaxy bias of ALMA-detected H-dropouts (red star) and its comparison to other populations, including the brightest LBGs ('Massive LBGs'; blue triangles) at $z \sim 4-5$ (ref. \cite{Harikane:2018}), massive passive galaxies ('Passive galaxies': purple squares) with $M_{*} > 10^{10.5} M_{\odot}$ at $z=2-3$ (ref. \cite{Hartley:2013}), local massive ellipticals with $L=2-4 L_{*}$ ('Ellipticals'; dark-red-shaded region) and clusters ('Clusters'; grey-shaded region). Error bars, 1$\sigma$ estimated from Poisson statistics. Filled dark-blue and light-blue triangles denote massive and more typical ($L_{*}$) LBGs with UV magnitudes of $M_{UV} \approx -22$ and  $M_{UV} \approx -20.5$, respectively. Dotted lines, the corresponding galaxy bias for fixed halo mass (labelled) at different redshifts\cite{MoH:2002}; dashed line, the evolutionary track\cite{Fakhouri:2010} for galaxies with the same galaxy bias as H-dropouts. The descendants of H-dropouts are consistent with massive ellipticals at $z \sim 2-3$ and today's most massive galaxies residing in massive groups and clusters.
\label{Fig:clustering}
}
\end{figure}

To put the H-dropouts in the context of the cosmic evolution of massive galaxies, we probe their clustering properties through their cross-correlation with $H$-detected galaxies at $3.5 < z < 5.5$ from the CANDELS survey in the same three fields (Extended Data Fig.~8, Methods).
The derived galaxy bias, that is, the relationship between the spatial distribution of galaxies and the underlying dark matter density field, for the H-dropouts is $b = 8.4\pm1.5$, corresponding to a dark matter halo mass of $M_{\rm h} \sim 10^{13\pm0.3} h^{-1} M_{\odot}$ at $z=4$ (Fig.~\ref{Fig:clustering}, Methods). This halo mass of H-dropouts is consistent with them being progenitors of the most massive quiescent galaxies at $z = 2-3$, as well as progenitors of today's ellipticals that reside in the central region of massive groups and clusters. 

The discovery and confirmation of these H-dropouts as massive galaxies at $z \approx 3-6$ alleviates greatly the tension between the small number of massive LBGs at $z > 3$ and the rapid emergence of massive (and quiescent) galaxies at $z \approx 2-3$. 
Assuming an average redshift of $z \approx 4$ and SFR $\approx 220~M_{\odot}$ yr$^{-1}$, these H-dropouts will grow in stellar mass by $1.3 \times 10^{11} M_{\odot}$ before $z \sim 3$. Their number density, $n \sim 2 \times 10^{-5}$ Mpc$^{-3}$, is also comparable to that of the most massive, quiescent galaxies at $z \sim 3$ with\cite{Straatman:2014} $M_{*} > 10^{11} M_{\odot}$. The early formation of such a large number of massive, dusty galaxies is unexpected with current semi-analytical models\cite{Henriques:2015}, which underestimates their density by one to two orders of magnitude (Fig.~\ref{Fig:SFRD}). Similarly, a deficit of such galaxies is also present in hydrodynamic simulations, which contain no such galaxies at $z > 3$ in mock deep fields ($\sim 23.5$ arcmin$^{2}$) from the Illustris Project~\cite{Snyder:2017}. Moreover, even  considering LBGs alone, the number of massive galaxies already appears too large when compared to the number of massive halos at $z >4$ predicted\cite{Steinhardt:2016} by our current understanding of galaxy evolution in the Lambda Cold Dark Matter (LCDM) framework. Together, this unexpected large abundance of massive galaxies in the early Universe suggests that our understanding of massive-galaxy formation may require substantial revision.  Spectroscopic follow-up of the whole population of H-dropouts would be key to providing further insights into this question, which calls for mid-infrared spectroscopy with James Webb Space Telescope in the near future.

\end{bibunit}

\clearpage

\setcounter{table}{0}

\setcounter{figure}{0}

\makeatletter
\renewcommand{\figurename}{Extended Data Fig.}
\renewcommand{\tablename}{Extended Data Table.}

\makeatother

\begin{methods}

\newbibstartnumber{30}
\begin{bibunit}[plainnat]

Here we give details of the multi-wavelength observations and the estimation of physical properties of sample galaxies. Throughout we adopt a Chabrier  initial mass function\cite{Chabrier:2003} and the concordance cosmology with $\Omega_{M} = 0.3$, $\Omega_{\Lambda}$ = 0.7, and $H_{0} =70$ km s$^{-1}$ Mpc$^{-1}$. All magnitudes are in the AB system.\\

\section{Observations}
\subsection{Selection of $H-$dropouts and incompleteness correction}

We have crossmatched the F160W-selected catalog from the three CANDELS fields (Table~\ref{TAB:fields})
with an IRAC 3.6 and 4.5 $\mu$m selected catalog\cite{Ashby:2013} from the SEDS survey. The SEDS survey
covers the three fields of H-dropouts to a depth of 26 AB mag (3$\sigma$) at
both 3.6 and 4.5 $\mu$m and is 80\% complete down to [4.5] $\sim$ 24 mag. We first matched sources with $[4.5] < 24$ mag in the
SEDS catalog to the F160W-selected catalog and identified those
without H-band counterparts within a 2" radius (corresponding roughly to the PSF size of IRAC 3.6 and 4.5 $\mu$m). This 4.5 $\mu$m
magnitude cut was applied to enable sufficient color range to
identify extremely red objects while keeping a complete
4.5 $\mu$m selected sample. We then visually inspected the IRAC
images and excluded sources whose flux is  contaminated
by bright neighbors as well as those falling on the edge of the
F160W image. With knowledge of their positions, some of these H-dropouts are
marginally detected in the F160W band but exhibit extended profiles
and are unidentifiable as real sources without that prior
knowledge. This left us 63 sources with 2 of them serendipitously detected in previous band-7 continuum observations with ALMA.

The criterion of no HST counterparts within 2" radius ensures a clean selection of H-dropouts with reliable constrains of  IRAC fluxes.
However, given the high density of HST sources in these deep fields, the chance probability of an IRAC-HST coincidence (with distance $<$ 2") is non-negligible  . This means that we may have missed some $H-$dropouts simply due to the presence of a random HST source falling within the 2" search radius of the IRAC source. To correct for this effect, we calculate the completeness of this selection approach, which is defined as at a given position the probability of finding zero galaxies in the 2" radius, 
$p(n=0) = {\rm exp}(- N * \pi *$ radius$^2$), with $N$ representing the surface density of HST sources. Averaging over the three CANDELS fields yields $N = 0.05$ arcsec$^{-2}$, implying $p(n=0) = 0.53$. This suggests that while our approach yields a clean selection of $H-$dropouts, roughly half of the true $H-$dropouts have been missed simply due to chance superposition of sources, which needs to be corrected. In fact, this completeness correction is consistent with recent findings from a blind ALMA survey, which reveals four H-dropouts (with $[4.5] < 24$) that were not picked up by our approach within an area of 1/3 of the GOODS-South filed~\cite{Franco:2018,Yamaguchi:2019}, in comparison to 12 sources selected by our approach in the whole GOODS-South field. Among these four sources, 3 of them have at least one HST counterpart within 2'' (with the remaining one  absent from our IRAC catalog, which is shallower than the one used in \cite{Franco:2018}), which is inconsistent with being the right counterpart of the ALMA emission based on the redshift and other physical properties. Albeit with small number statistics, this implies a completeness of our searching approach of $\sim 57\%$, consistent with our estimated value.   
In addition to this correction, we need to also correct for the incompleteness of the IRAC imaging from the SEDS survey, which ranges from 93\% at [4.5] =22 to 75\% -80\% at $[4.5] = 24$ in the three fields. Combining the two corrections, a factor of 2 to 2.4 has been applied to the number density (including also star formation rate and stellar mass density) of H-dropouts depending on their IRAC fluxes.

\subsection{Multiwavelength photometry}

In each field, we gathered mosaics in a large number of bands, including 
all the images used to build the 3DHST \cite{Skelton:2014} and ZFOURGE 
\cite{Straatman:2016} catalogs. All our galaxies therefore had rich and 
deep photometry from the UV to the NIR, reaching typical $5\sigma$ 
depths (AB) of $27$ in {\it u} to {\it i}, $26$ in {\it z} to {\it H}, 
and $25$ in {\it Ks}. We provide the full detail of the used mosaics below.

For GOODS-South, we used VLT/VIMOS images in the {\it U} and {\it R} 
bands \cite{Nonino:2009}, ESO/WFI images in the {\it U}, {\it U38}, {\it 
B}, {\it V}, {\it R}, {\it I} bands from GaBoDS \cite{Hildebrandt:2006}, CTIO/MOSAIC image in the {\it z} band from MUSYC \cite{Gawiser:2006}, 
Subaru images in 15 medium bands from MUSYC \cite{Cardamone:2010}, 
{\it Hubble} images in the F395W, F606W, F775W, F8514W, F850LP, F105W, 
F125W, F160W bands from GOODS and CANDELS programs 
\cite{Giavalisco:2004,Grogin:2011,Koekemoer:2011}, VLT/ISAAC images in the {\it J}, 
{\it H}, {\it Ks} bands \cite{Retzlaff:2010}, CFHT/WIRCam images in the 
{\it J} and {\it Ks} bands from TENIS \cite{Hsieh:2012}, 
Magellan/FOURSTAR images in the {\it J1}, {\it J2}, {\it J3}, {\it Hs}, 
{\it Hl}, {\it Ks} bands from ZFOURGE \cite{Straatman:2016}, a 
VLT/HawK-I image in the {\it Ks} band from HUGS \cite{Fontana:2014}, and 
{\it Spitzer} IRAC images from SEDS \cite{Ashby:2013}.

For UDS, we used a CFHT/Megacam image in the {\it u} band produced by 
the 3DHST team \cite{Skelton:2014}, Subaru images in the {\it B}, {\it 
V}, {\it R}, {\it i}, {\it z} bands \cite{Furusawa:2008}, {\it Hubble} 
images in the F606W, F814W, F125W, F140W, F160W bands from the CANDELS 
and 3DHST programs \cite{Koekemoer:2011,Momcheva:2016}, UKIRT/WFCAM 
images in the {\it J}, {\it H}, {\it K} bands from UKIDSS 
\cite{Lawrence:2007}, Magellan/FOURSTAR images in the {\it J1}, {\it 
J2}, {\it J3}, {\it Hs}, {\it Hl}, {\it Ks} bands from ZFOURGE 
\cite{Straatman:2016}, VLT/HawK-I images in the {\it Y} and {\it Ks} 
bands from HUGS \cite{Fontana:2014}, and {\it Spitzer} IRAC images from 
SEDS \cite{Ashby:2013} and SpUDS (PI: J.~Dunlop).

For COSMOS, we used CFHT/Megacam images in the {\it u} and {\it i} bands 
from CFHTLS \cite{Cuillandre:2012}, Subaru images in the {\it B}, {\it 
g}, {\it V}, {\it r}, {\it i}, {\it z} bands as well as 10 medium bands 
\cite{Taniguchi:2007}, {\it Hubble} images in the F606W, F814W, F125W, 
F140W, F160W bands from the CANDELS and 3DHST programs 
\cite{Koekemoer:2011,Momcheva:2016}, CFHT/WIRCam images in the {\it H} 
and {\it Ks} bands \cite{mccracken:2010}, Magellan/FOURSTAR images in 
the {\it J1}, {\it J2}, {\it J3}, {\it Hs}, {\it Hl}, {\it Ks} bands 
from ZFOURGE \cite{Straatman:2016}, VISTA/VIRCAM images in the {\it Y}, 
{\it J}, {\it H}, {\it Ks} from UltraVISTA DR3 \cite{Mccracken:2012}, 
and {\it Spitzer} IRAC images from SEDS \cite{Ashby:2013} and S-COSMOS 
\cite{Sanders:2007}.

The photometry was obtained with a procedure very similar to that previously used in deep surveys \cite{Straatman:2016,Momcheva:2016}, which we summarize here. Fluxes in UV-to-NIR were extracted on re-gridded and PSF-matched images in fixed apertures of 2$\arcsec$ diameter. Because of the broader PSF in Spitzer images, fluxes in the IRAC bands were extracted separately, with a 3" aperture and without PSF matching. The asymmetric IRAC PSF was rotated to match the telescope roll angle for each field. Prior to extracting the fluxes, all the neighboring sources within a 10" radius were subtracted from the images. This was done by identifying the sources from a stacked detection image, and using the HST F160W profile of each source as a model. These models were convolved by the PSF of each image, where they were fit simultaneously using a linear solver. Most often the dropouts were not found in the stacked detection image, and were therefore modeled as point-sources at the coordinates of their IRAC centroid during the de-blending stage. Once the flux was extracted, additional "sky" apertures were placed randomly around each dropout. The median flux in these sky apertures was subtracted from the dropout's flux, to eliminate any remaining background signal, while the standard deviation of these fluxes was used as flux uncertainty. Lastly, fluxes and uncertainties were aperture-corrected using the matched PSF's light curve, assuming point-like morphology.

\subsection{ALMA observation and data reduction}

Our ALMA band-7 continuum observations of H-dropouts are performed during January and July  2016. The observations were centered on the IRAC positions with a spectral setup placed around a central frequency of 343.5 GHz. While we asked 0.7$\arcsec$-resolution observation for all the three fields, only the CANDLES-COSMOS field was observed as requested, yielding a synthesis beam of $0.6 \times 1\arcsec$. The other two fields were observed at 0.2-0.3$\arcsec$ resolution. The integration time is roughly 1.8 mins per object with a total observing time of $\sim$2 h. We reduced the data using the CASA pipeline (version 4.3.1). To reach an homogeneous angular resolution, we tapered the baselines for these two fields to an angular resolution of 0.6$\arcsec$. This resolution corresponds to $\sim$ 4 kpc at $z =4$, compared to typical sizes of $\sim$2 kpc for SMGs~\cite{Hodge:2016}.

We measured the total flux of all our targets directly in the $(u, v)$ plane using the uvmodelfit procedure from the CASA pipeline. The sources were  modeled with a circular Gaussian profile of variable total flux, centroid, width, axis ratio and position angle. 39 H-dropouts are detected at S/N$>4$ with $S_{870 \mu m} > 0.6$ mJy, including two galaxies that were serendipitously detected in a previous ALMA program\cite{Schreiber:2016} targeting $H$-detected $z\sim4$ galaxies, which has reached similar depth as this observation. The positions of the 870 $\mu$m emission as measured from ALMA are in good agreement with IRAC, with $\Delta$RA=0.081$\pm0.128\arcsec$ and $\Delta$DEC= -0.13$\pm0.16\arcsec$.

\subsection{SCUBA-2 450 $\mu$m and VLA  observations }

One of the three $H$-dropout fields, CANDELS-COSMOS, is covered by deep SCUBA-2 450 $\mu$m and 870 $\mu$m observations from the STUDIES survey\cite{WangW:2017}. Previous observations with JCMT/SCUBA-2 at the same region\cite{Casey:2013, Geach:2013,Geach:2017} have also been combined to produce an extremely deep 450 $\mu$m image and a confusion-limited 850 $\mu$m image. The instrumental noises at 450 $\mu$m and 850 $\mu$m at the deepest regions reach $\sim$0.65 mJy  and $\sim 0.1$ mJy, respectively.

The SCUBA2-450 $\mu$m and -850 $\mu$m fluxes for H-dropouts are measured at the position of the IRAC 3.6 and 4.5 $\mu$m emission with the prior-based PSF-fitting code FASTPHOT\cite{Bethermin:2010b}.  
We further restrict all the extracted fluxes to be positive with bounded value least-square minimization. 
During the fit we have included all the MIPS 24 $\mu$m- and VLA  detections as priors to perform source extraction. The VLA  3 GHz observation in COSMOS\cite{Smolcic:2017} reaches a rms of 2.3 $\mu$Jy/beam at an angular resolution of 0.75$\arcsec$, which is deep enough to put useful constraints on their redshifts. The flux measurement for H-dropouts  in the far-IR suffers minimum source confusion due to our selection criterion (no close neighbours within a 2$\arcsec$ radius). A comparison of 870 $\mu$m fluxes measured by  ALMA and SCUBA-2 yields excellent agreement with a median value of $S_{\rm ALMA}/S_{\rm SCUBA-2} = 1.05$.

\subsection{ X-SHOOTER spectra}
In the COSMOS field, deep medium band images in the optical were obtained with the Subaru telescope\cite{Taniguchi:2007}. We visually inspected these images at the location of each dropout in our sample and found three galaxies with flux excesses in one of these images, with a significance above 4 sigma. Examples are shown on Extended Data Figure~\ref{Fig:X-shooter}. Such flux excess can be interpreted as coming from a bright emission line\cite{Sobral:2018}. For these three dropouts, the line could be identified as Ly$\alpha$ at $z=5.0$, $3.2$ and $4.1$, respectively. Even though H-dropouts are typically very obscured, Ly$\alpha$ may still be detected through un-obscured sight lines, or by scattering\cite{Finkelstein:2008}. Judging from the spatial offsets of about $1\arcsec$ we observed between this optical flux excess and the {\it Spitzer}--IRAC or ALMA emission, scattering appears to be the most plausible explanation.

We thus followed up these objects with VLT/X-SHOOTER to confirm the presence of an emission line. Each dropout was observed in May 2018 in the UVB and VIS arms for 50 minutes in stare mode (no nodding), split in three exposures. The 2D spectra were reduced using the standard pipeline, and 1D spectra were produced by fitting a Gaussian profile to each spectral slice. Uncertainties were controlled by computing the standard deviation of spectral elements in regions without sky lines; we found that the 1D uncertainty spectrum had to be rescaled upwards by a factor 1.27 to match the observed noise. 

We then searched for emission lines in the spectra, considering only the wavelength range covered by the Subaru medium band in which the flux excess was previously identified. The result of this search is displayed on Extended Data Figure~\ref{Fig:X-shooter}. We found a 10$\sigma$ detection at 0.498$\mu$m for the dropout 32932, corresponding to $z_{\rm spec} = 3.0971\pm0.0002$, and a more marginal but still significant 4.3$\sigma$ detection at 0.739$\mu$m for the dropout 25363, corresponding to $z_{\rm spec} = 5.113^{+0.001}_{-0.005}$. Because our search space is tightly limited by the Subaru passband, the latter only has a 0.4\% chance of being spurious, and we therefore consider it a reliable detection. The third dropout showed no significant line emission above $2\sigma$. 

\subsection{Lyman-break galaxy selection}
In order to compare the properties of $H-$dropouts and LBGs\cite{Steidel:1996}, we have selected LBGs using the ZFOURGE catalogs in the same three CANDELS fields\cite{Straatman:2016}.  The advantage of the ZFOURGE catalog is that it is essentially a $K{_s}-$band selected catalog, for which the deep $K_{s}$ band data provides critical constraints on the redshift and stellar masses estimates at $z > 4$. We select our $z=4-6$ LBG galaxy sample using the selection criterion in \cite{Bouwens:2015}. Due to the lack of B-band data from HST, the $z\sim4$ LBG sample is only limited to GOODS-South field while the $z\sim5$ and $z\sim6$ LBG sample include galaxies from all the three fields. To enable a clean selection of galaxies with reliable flux density measurements, we have further limited the selection to galaxies with $use=1$  as recommended\cite{Straatman:2016}. This reduces the effective area to 132.2, 139.2, and 135.6 arcmin$^{2}$ for GOODS-South, COSMOS, and UDS, respectively. To identify total SFRD from massive LBGs with $M_{*} > 10^{10.3} M_{\odot}$, we utilized the latest determination of the UV luminosity function at $z \sim 4-6$\cite{Ono:2018}. Taking into account variations in the $M_{*}-M_{UV}$ relation, this mass cut  corresponds to $M_{UV} < [-21.55, -22.04, -22.27]$ at $z=[4, 5, 6]$, respectively. We then derive the dust-corrected SFR for these brightest UV-selected galaxies following the approach in ref.\cite{Bouwens:2012}.

\section{Determination of Physical Properties}
\subsection{Stacked UV-to-NIR SEDs}
To produce the stacked UV-to-NIR SEDs, we took the fluxes of each galaxy in our photometric catalog and normalized them by their respective IRAC $4.5\,\mu$m flux. We then computed the mean flux in each band, using inverse variance weighting, and finally multiplied the resulting stacked fluxes by the average $4.5\,\mu$m flux of the stacked sample. In the stack, we combined bands that have similar effective wavelengths, even though the true passbands could be slightly different; for example we stacked together all the $K$s bands from UKIDSS, UltraVISTA, FOURSTAR, WIRCam, and ISAAC into a single $K$s band. The uncertainties on the stacked fluxes were derived by formally combining the uncertainties of each stacked galaxy. We note that, since we obtained our photometry using fixed-size apertures, this method is strictly equivalent to stacking the images.

\subsection{Photometric redshift and stellar mass determination}

Using the aforementioned multiwavelength photometry, including bands with formal non-detections, photometric redshifts were computed with EAzY \cite{Brammer:2008} using the full set of template SEDs, namely, including the old-and-dusty template and the extreme emission line template. The prior on the observed magnitudes was not used. Using these redshifts, we then ran FAST \cite{Kriek:2009a} to estimate the stellar masses. We assumed a delayed-exponentially-declining star-formation history, with a range of age and exponential timescale. Dust attenuation was modeled with the \cite{Calzetti:2001} prescription, allowing $A_V$ up to 6 magnitudes. Metallicity was fixed to solar during the fitting. We also used 
the infrared luminosities inferred from the ALMA fluxes to further 
constrain the fits. This was implemented as follows. From the stacked 
Herschel SED (see Figure~\ref{Fig:stacked_IR}), we measured the mean dust temperature of 
our sample: $T_{\rm dust}=36.7 \pm 2.1\,{\rm K}$. Based on Herschel and 
ALMA observations of $z>2$ galaxies \cite{Schreiber:2018a}, we 
expect a typical scatter of $5\,{\rm K}$ around the average temperature 
at any given redshift. Assuming this distribution of temperatures holds 
for the dropouts, we generated probability distributions for $L_{\rm 
IR}$ using a Monte Carlo procedure: the measured ALMA flux was randomly 
perturbed with Gaussian noise of amplitude set by the flux uncertainty, 
and the dust temperature was drawn from a Gaussian distribution centered 
on $36.7\,{\rm K}$ and with a width of $5\,{\rm K}$; the resulting dust 
SED was then used to extrapolate $L_{\rm IR}$ from the ALMA measurement. 
For galaxies whose ALMA flux has $S/N<2$, the resulting probability 
distribution of $L_{\rm IR}$ was close to Gaussian, while for the 
detections the probability distribution was close to log-normal. We 
modeled these two regimes accordingly in the fit, by assuming either 
Gaussian noise on $L_{\rm IR}$ or $\log_{10}(L_{\rm IR})$, respectively. 
The observed infrared luminosity was then compared to the modeled value, 
which we computed as the difference of bolometric luminosity before 
after applying dust attenuation. This resulted in an additional 
contribution to the $\chi^2$, which was then used for standard model 
selection.

Uncertainties on the photometric redshifts were derived from the 16th 
and 84th percentiles of the probability distribution produced by EAzY. 
This accounts for uncertainty in the photometry as well as on the model 
galaxy templates. Uncertainties on the derived physical parameters, 
including the stellar mass, were derived using Monte Carlo simulations, 
where the observed photometry was randomly perturbed with Gaussian noise 
of amplitude determined by the estimated photometric uncertainties. This 
was repeated 200 times. The error bars on physical parameters were then 
derived from the 16th and 84th percentiles of the distribution of the %
values obtained in the Monte Carlo simulations. For each fit, the 
redshift was left free to vary within the 68\% confidence interval 
reported by the photometric redshift code. Therefore the resulting error 
bars account for uncertainties on the photometry and on the redshift.

\subsection{Clustering measurements}

Since the number of H-dropouts is small, we calculate two-point angular cross-correlation function (CCF) with a much larger population of galaxies 
sharing the same cosmic volume (redshifts) in order to enhance the statistics. Specifically we select all the galaxies with $3.5 < z < 5.5$ from the {\it H}-selected 
catalog in the same three CANDELS fields (``the galaxy sample'', hereafter), and then calculate CCF using the estimator as follows \cite{Landy:1993}: 
\begin{equation}
\omega (\theta) = \frac{HG(\theta) - HR(\theta) - GR(\theta) + RR(\theta)}{RR(\theta)}
\end{equation}
where $HG$, $HR$, $GR$, $RR$ are Hdropout-galaxy, Hdropout-random, galaxy-random and random-random pair counts respectively. The random galaxy sample are created within the same CANDELS footprint as the H-dropouts (we exclude HUDF in the GOODS-S field because of its much deeper integration than other regions).
The uncertainties of CCF are estimated as:
\begin{equation}
\Delta \omega (\theta) = \frac{1 + \omega (\theta)}{\sqrt{HG (\theta)}}.
\end{equation}
We then fit the derived CCF with a power-law model:
\begin{equation}
\omega (\theta) = A_\omega \theta^{-\beta} - IC,
\end{equation}
where $A_\omega$ is the correlation amplitude and $\beta$ is the power-law index fixed to 0.8 and $IC$ is the integral constraint. Integral constraints is an offset due to the clustering measurement over the limited area and is calculated by
\begin{equation}
IC = \frac{\sum RR (\theta) A_\omega \theta^{-\beta}}{\sum RR (\theta)}.
\end{equation}
The derived correlation amplitude can be converted to three-dimensional correlation length $r_0$ by Limber equation\cite{Limber:1953} modified by \cite{Croom:1999} for the cross-correlation.

The correlation length is related to galaxy bias $b$, such that
\begin{equation}
\sigma_{8,\mathrm{gal}}^2 = \frac{72}{(3-\gamma)(4-\gamma)(6-\gamma)2^\gamma} \left(\frac{r_0}{8\, h^{-1}\mathrm{Mpc}} \right)^\gamma
\end{equation}
and 
\begin{equation}
b = \frac{\sigma_{8,\mathrm{gal}}}{\sigma_8 (z)},
\end{equation}
where $\sigma_{8,\mathrm{gal}}$ is a galaxy fluctuation,  $\gamma = 1 + \beta$, and $\sigma_8 (z)$ is a matter fluctuation\cite{Peebles:1993}.
The halo mass is then derived from the estimated galaxy bias\cite{MoH:2002}.\\

\end{bibunit}




\end{methods}

\begin{extendeddata}

\begin{table*}[htp]
\caption{\textbf{Survey depths for each field} \label{TAB:fields}}
\begin{center}
\begin{tabular}{lccccc}
    \hline
    \hline \\[-2.5mm]
    Field & Area & WFC3/F160W ($5\sigma$) & H-dropouts & ALMA-detected & ALMA-undetected \\
          &       ${\rm arcmin}^2$       &   &    ($[4.5] < 24$)      &         ($S_{870 \mu m} > 0.6$ mJy) &   ($S_{870 \mu m} < 0.6$ mJy) \\
    \hline \\[-2.5mm]
    CANDELS-GDS     & 184 & $H < 27.4$--$29.7$ & 12 & 10 & 2  \\
    CANDELS-UDS    & 202 & $H < 27.1$--$27.6$ & 33 & 14 & 19 \\
    CANDELS-COSMOS & 208 & $H < 27.4$--$27.8$ & 18 & 15 & 3 \\
    \hline
\end{tabular}
\end{center}
\end{table*}

\begin{table*}[htp]
\begin{center}
\caption{\textbf{Physical properties of H-dropouts} \label{Tab:measurements} }

\begin{tabular}{lcccccc}
    \hline
    \hline \\[-2.5mm]
    ID & R.A.  & Decl.  & [4.5]  & $S_{870\mu m}$ & $z$   & Log $M_{*}$   \\
& (J2000) & (J2000) & & (mJy) & & $M_{\odot}$ \\
\hline 
GDS-25526 & 03:32:47.97 & -27:54:16.4 &  22.05 &      8.34$\pm$0.18  &  4.74$^{+0.28}_{-0.30}$  	  &   10.84$^{+0.05}_{-0.17}$\\
GDS-27571 & 03:32:30.62 & -27:42:24.3 &  22.44 &      0.82$\pm$0.16  &  4.64$^{+0.19}_{-1.64}$  	  &   10.89$^{+0.04}_{-0.48}$\\
GDS-40613 & 03:32:11.44 & -27:52:07.1 &  23.17 &      1.83$\pm$0.18  &  3.04$^{+0.22}_{-0.35}$  	  &   10.64$^{+0.02}_{-0.22}$\\
GDS-43215 & 03:32:20.34 & -27:42:28.8 &  23.02 &      1.52$\pm$0.16  &  2.91$^{+0.19}_{-0.22}$  	  &   10.25$^{+0.03}_{-0.53}$\\
GDS-44539 & 03:32:28.59 & -27:48:50.2 &  23.41 &      0.69$\pm$0.10  &  4.22$^{+0.77}_{-0.67}$  	  &   10.94$^{+0.03}_{-0.31}$\\
GDS-47375 & 03:32:14.62 & -27:43:06.0 &  23.48 &      1.89$\pm$0.12  &  3.60$^{+0.66}_{-0.66}$  	  &   10.27$^{+0.13}_{-0.12}$\\
GDS-48764 & 03:32:32.31 & -27:54:26.9 &  23.31 &      2.54$\pm$0.42  &  5.16$^{+3.08}_{-1.73}$  	  &   10.38$^{+0.51}_{-0.03}$\\
GDS-48885 & 03:32:47.17 & -27:45:25.1 &  23.57 &      0.87$\pm$0.11  &  4.62$^{+0.19}_{-0.17}$  	  &   10.56$^{+0.02}_{-0.08}$\\
GDS-49094 & 03:32:31.85 & -27:43:12.7 &  23.59 &      0.89$\pm$0.13  &  3.69$^{+0.29}_{-0.28}$  	  &   10.29$^{+0.10}_{-0.05}$\\
GDS-52734 & 03:32:10.10 & -27:50:33.1 &  24.06 &      1.41$\pm$0.15  &  5.13$^{+1.72}_{-1.18}$  	  &   10.71$^{+0.17}_{-0.21}$\\
GDS-54513 & 03:32:04.99 & -27:41:56.5 &  23.71 &    	$<$0.6       & 4.33$^{+0.32}_{-0.37}$  	  	  &   10.30$^{+0.00}_{-0.37}$\\
GDS-58560 & 03:32:40.11 & -27:42:55.3 &  23.85 &    	$<$0.6       & 5.35$^{+2.33}_{-2.18}$  	      &   10.58$^{+0.39}_{-0.43}$\\   
COS-16199 & 10:00:25.41 & +02:25:43.9 &  21.96 &      3.91$\pm$0.09  &  6.54$^{+1.43}_{-1.54}$  	  &   10.90$^{+0.14}_{-0.27}$\\
COS-19762 & 10:00:15.89 & +02:24:45.9 &  22.94 &      4.35$\pm$0.10  &  3.52$^{+5.36}_{-0.19}$  	  &   10.79$^{+0.88}_{-0.08}$\\
COS-23718 & 10:00:28.95 & +02:25:05.3 &  22.82 &      2.25$\pm$0.10  &  5.77$^{+0.80}_{-0.88}$  	  &   11.02$^{+0.13}_{-0.24}$\\
COS-23913 & 10:00:23.03 & +02:21:55.0 &  22.87 &      1.63$\pm$0.09  &  3.65$^{+0.35}_{-0.29}$  	  &   10.65$^{+0.10}_{-0.11}$\\
COS-24466 & 10:00:38.07 & +02:28:06.2 &  23.22 &      1.38$\pm$0.09  &  3.35$^{+0.39}_{-0.38}$  	  &   10.37$^{+0.34}_{-0.26}$\\
COS-25270 & 10:00:23.62 & +02:13:57.4 &  23.47 &      $<$0.6         &  3.78$^{+0.55}_{-0.56}$  	  &   10.48$^{+0.04}_{-0.20}$\\
COS-25363 & 10:00:26.68 & +02:31:26.2 &  23.15 &      3.0 $\pm$0.2   &  5.113$^{+0.001}_{-0.005}$     &   10.52$^{+0.09}_{-0.19}$\\
COS-25881 & 10:00:27.03 & +02:24:24.0 &  22.96 &      1.30$\pm$0.10  &  6.58$^{+1.43}_{-1.38}$  	  &   11.12$^{+0.17}_{-0.30}$\\
COS-27285 & 10:00:27.79 & +02:25:52.2 &  23.74 &      1.58$\pm$0.09  &  4.32$^{+0.23}_{-0.22}$  	  &   10.31$^{+0.14}_{-0.06}$\\
COS-27392 & 10:00:27.98 & +02:25:29.7 &  23.42 &      $<$0.6         &  3.61$^{+0.51}_{-0.49}$  	  &   10.38$^{+0.11}_{-0.08}$\\
COS-30182 & 10:00:14.70 & +02:28:01.7 &  23.08 &      1.59$\pm$0.09  &  6.37$^{+1.16}_{-2.15}$  	  &   10.94$^{+0.22}_{-0.28}$\\
COS-30614 & 10:00:14.69 & +02:30:04.6 &  23.39 &      0.85$\pm$0.10  &  3.97$^{+0.19}_{-0.29}$  	  &   10.22$^{+0.04}_{-0.14}$\\
COS-31278 & 10:00:26.09 & +02:12:31.6 &  23.36 &      0.93$\pm$0.09  &  3.37$^{+0.43}_{-0.35}$  	  &   10.14$^{+0.13}_{-0.15}$\\
COS-31483 & 10:00:46.50 & +02:23:09.1 &  23.89 &      0.70$\pm$0.11  &  2.97$^{+0.40}_{-0.41}$  	  &    9.91$^{+0.13}_{-0.13}$\\
COS-31661 & 10:00:41.83 & +02:25:47.0 &  23.28 &      2.88$\pm$0.12  &  3.72$^{+0.17}_{-0.19}$  	  &   10.36$^{+0.05}_{-0.20}$\\
COS-32409 & 10:00:15.84 & +02:23:04.0 &  23.66 &      0.70$\pm$0.11  &  3.91$^{+1.66}_{-1.16}$  	  &    9.87$^{+0.48}_{-0.06}$\\
COS-32932 & 10:00:22.44 & +02:23:41.1 &  23.22 &      $<$0.6         &  3.0971$^{+0.0002}_{-0.001}$   &    9.96$^{+0.15}_{-0.11}$\\
COS-34487 & 10:00:35.34 & +02:28:26.7 &  23.36 &      4.3 $\pm$0.15  &  3.15$^{+0.52}_{-0.60}$        &   10.18$^{+0.26}_{-0.20}$\\
UDS-24945 & 02:16:59.77 & -05:11:52.8 &  22.16 &      $<$0.6         &  3.50$^{+0.76}_{-0.61}$        &   10.76$^{+0.22}_{-0.11}$\\
UDS-29006 & 02:17:05.52 & -05:08:45.8 &  22.71 &      $<$0.6         &  3.79$^{+0.54}_{-0.53}$        &   10.46$^{+0.12}_{-0.21}$\\
UDS-31037 & 02:18:07.67 & -05:13:26.8 &  22.58 &      1.83$\pm$0.13  &  3.62$^{+0.70}_{-0.65}$        &   10.49$^{+0.26}_{-0.07}$\\
UDS-31072 & 02:17:43.32 & -05:11:57.4 &  22.29 &      3.63$\pm$0.25  &  4.20$^{+4.17}_{-0.44}$        &   11.31$^{+0.33}_{-0.09}$\\
UDS-31959 & 02:18:11.36 & -05:16:23.7 &  22.98 &      $<$0.6         &  3.88$^{+0.68}_{-0.73}$        &   10.22$^{+0.19}_{-0.14}$\\
UDS-34637 & 02:18:05.80 & -05:11:23.1 &  22.89 &      1.07$\pm$0.37  &  2.84$^{+0.54}_{-0.44}$        &   10.33$^{+0.16}_{-0.17}$\\
UDS-37344 & 02:18:02.86 & -05:15:05.4 &  23.36 &      $<$0.6         &  2.89$^{+1.43}_{-1.48}$        &   10.18$^{+0.08}_{-0.63}$\\
UDS-37423 & 02:18:10.02 & -05:11:31.5 &  23.07 &      1.12$\pm$0.15  &  7.47$^{+0.63}_{-0.64}$        &   11.17$^{+0.20}_{-0.08}$\\
UDS-37560 & 02:17:03.44 & -05:15:51.3 &  22.72 &      4.40$\pm$0.14  &  3.95$^{+0.35}_{-0.34}$        &   10.35$^{+0.22}_{-0.17}$\\
UDS-37649 & 02:17:36.95 & -05:16:07.3 &  23.14 &      1.25$\pm$0.31  &  2.82$^{+0.44}_{-0.48}$        &   10.31$^{+0.11}_{-0.07}$\\
UDS-40772 & 02:17:36.56 & -05:12:52.0 &  23.19 &      1.96$\pm$0.31  &  4.00$^{+0.98}_{-0.98}$        &   10.87$^{+0.00}_{-0.40}$\\
UDS-41502 & 02:17:18.03 & -05:11:03.9 &  23.23 &      1.59$\pm$0.20  &  3.73$^{+0.44}_{-0.43}$        &   10.59$^{+0.09}_{-0.33}$\\
UDS-41525 & 02:16:59.59 & -05:14:15.4 &  23.34 &      $<$0.6         &  6.13$^{+1.08}_{-1.01}$        &   11.17$^{+0.15}_{-0.16}$\\
UDS-41773 & 02:18:07.02 & -05:09:18.1 &  23.11 &      1.54$\pm$0.15  &  3.52$^{+1.35}_{-0.84}$        &   10.20$^{+0.56}_{-0.24}$\\
UDS-42280 & 02:18:11.16 & -05:10:27.1 &  23.88 &      1.16$\pm$0.37  &  4.21$^{+0.18}_{-0.17}$        &   10.06$^{+0.09}_{-0.16}$\\
UDS-42875 & 02:18:21.15 & -05:09:42.5 &  23.04 &      1.74$\pm$0.15  &  7.20$^{+1.20}_{-1.83}$        &   11.75$^{+0.05}_{-0.34}$\\
UDS-43941 & 02:17:43.65 & -05:14:23.9 &  23.46 &      1.91$\pm$0.11  &  3.39$^{+0.23}_{-0.23}$        &   10.17$^{+0.02}_{-0.10}$\\
UDS-44515 & 02:18:20.89 & -05:11:11.1 &  23.48 &      $<$0.6         &  4.20$^{+1.35}_{-1.35}$        &   10.59$^{+0.07}_{-0.27}$\\
UDS-44594 & 02:17:20.20 & -05:11:55.4 &  23.79 &      0.66$\pm$0.14  &  4.44$^{+0.52}_{-0.39}$        &   10.60$^{+0.04}_{-0.17}$\\
UDS-45868 & 02:18:15.00 & -05:10:02.7 &  23.61 &      $<$0.6         &  3.66$^{+0.88}_{-0.88}$        &   10.45$^{+0.13}_{-0.13}$\\
UDS-46241 & 02:17:58.31 & -05:15:00.3 &  23.44 &      $<$0.6         &  2.18$^{+0.78}_{-0.53}$        &    9.72$^{+0.22}_{-0.38}$\\
UDS-46513 & 02:18:17.87 & -05:11:53.9 &  23.84 &      0.58$\pm$0.11  &  3.54$^{+0.95}_{-0.93}$        &   10.31$^{+0.11}_{-0.19}$\\
UDS-46648 & 02:17:08.17 & -05:15:37.8 &  23.52 &      2.27$\pm$0.20  &  6.88$^{+1.67}_{-1.74}$        &   11.37$^{+0.30}_{-0.43}$\\
UDS-46693 & 02:17:59.07 & -05:09:37.5 &  23.47 &      $<$0.6         &  3.56$^{+0.37}_{-0.26}$        &   10.24$^{+0.05}_{-0.29}$\\
UDS-48514 & 02:17:29.83 & -05:14:23.5 &  23.65 &      $<$0.6         &  2.59$^{+0.93}_{-0.83}$        &   10.04$^{+0.03}_{-0.42}$\\
UDS-49119 & 02:17:07.14 & -05:12:54.0 &  23.77 &      1.34$\pm$0.30  &  4.60$^{+1.94}_{-1.83}$        &   10.65$^{+0.01}_{-0.53}$\\
UDS-49199 & 02:18:21.40 & -05:11:46.3 &  23.6  &      $<$0.6     	 &  3.96$^{+0.84}_{-0.97}$        &   10.66$^{+0.04}_{-0.24}$\\
UDS-49594 & 02:18:01.13 & -05:13:45.7 &  23.65 &      $<$0.6         &  3.77$^{+0.48}_{-0.55}$        &    9.79$^{+0.03}_{-0.26}$\\
UDS-49784 & 02:17:37.48 & -05:09:47.7 &  23.6  &      $<$0.6         &  3.95$^{+1.63}_{-2.21}$        &    9.91$^{+0.41}_{-0.77}$\\
UDS-51119 & 02:17:58.29 & -05:11:44.7 &  23.75 &      $<$0.6         &  3.45$^{+1.16}_{-1.16}$        &   10.25$^{+0.12}_{-0.13}$\\
UDS-52324 & 02:17:06.27 & -05:09:48.3 &  23.49 &      2.69$\pm$0.16  &  4.95$^{+1.74}_{-1.61}$        &   10.62$^{+0.12}_{-0.26}$\\
UDS-54074 & 02:17:09.70 & -05:15:11.3 &  23.75 &      0.81$\pm$0.10  &  5.89$^{+1.90}_{-2.37}$        &   10.65$^{+0.17}_{-0.35}$\\
UDS-63094 & 02:17:53.06 & -05:11:25.5 &  23.98 &      $<$0.6         &  5.08$^{+0.24}_{-0.26}$        &   10.00$^{+0.08}_{-0.27}$\\
\end{tabular}
\end{center}
\end{table*}



\begin{figure*}[!htb]
\centering
\includegraphics[width=0.9\textwidth]{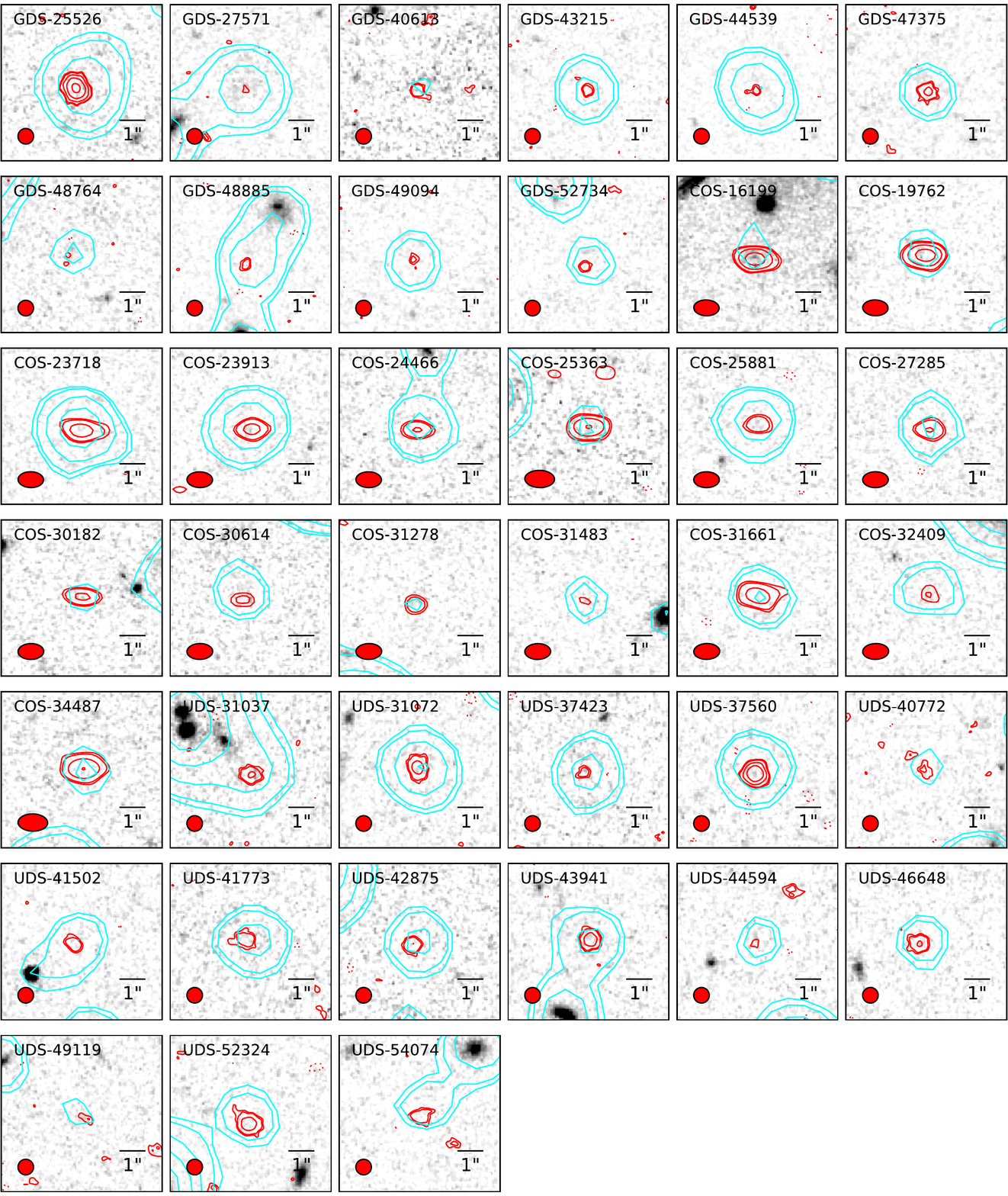}
\caption{\textbf{NIR and ALMA submillimeter-wavelength images of the ALMA-detected H-dropouts.} Images are 6\arcsec $\times$ 6\arcsec, centered at the centroid of the IRAC 4.5 $\mu$m emission. The greyscale images are F160W-band ({\it H}-band) exposures from the Hubble Space Telescope Wide Field Camera 3 (HST/WFC3). The red solid contours are ALMA  870 $\mu$m imaging, with contour levels starting at 3$\sigma$ and increasing as 4$\sigma$, 8$\sigma$, 16$\sigma$, 32$\sigma$, and 64$\sigma$. Negative contours at the same significances are shown with red dashed lines. The cyan contours are 4.5-$\mu$m emission, starting at 2$\sigma$ and increasing as 3$\sigma$, 4$\sigma$, 8$\sigma$ and 16$\sigma$. The exposure times for HST/WFC3 and ALMA imaging are roughly 2 h and 2 min per object, respectively. Although these H-dropouts are not detected in the deep F160W imaging (mag$_{\it H} \gtrsim 27$), they are significantly detected with ALMA within a short integration time.
\label{Fig:stamps_det}
}
\end{figure*}

\begin{figure*}[!htb]
\centering
\includegraphics[width=1\textwidth]{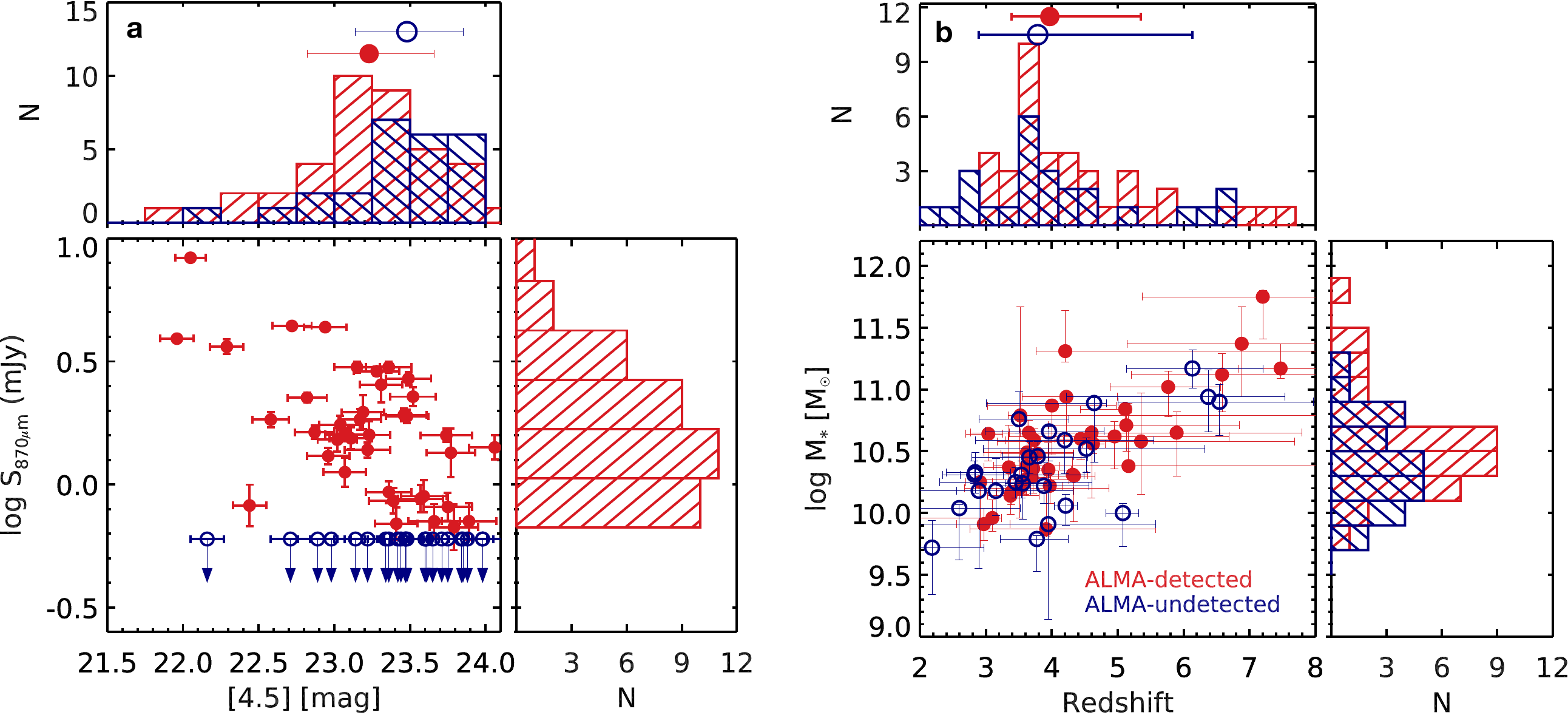}

\caption{\textbf{Physical properties of ALMA-detected and ALMA-undetected $H-$dropouts.} The ALMA-detected and un-detected $H-$dropouts are shown in  blue and red, respectively. {\bf a:} Main panel, the $870 \mu$m fluxes of ALMA-undetected H-dropouts are shown by their upper limits, $S_{870 \mu m} < 0.6$ mJy (4$\sigma$). The ALMA-undetected H-dropouts tend to have slightly fainter 4.5 $\mu$m magnitudes, with a median value of [4.5]$_{\rm median}$ = 23.5  compared to [4.5]$_{\rm median}$ = 23.2 for ALMA-detected ones. The error bars for ALMA-detected H-dropouts denote their 1$\sigma$ measurement error, while for ALMA-undetected H-dropouts their 4$\sigma$ upper limits are shown. Top panel, histogram showing the distribution of the 4.5-$\mu$m magnitudes of H-dropouts. The filled and open circles and their error bars denote the median 4.5-$\mu$m magnitude as well as the 16th and 84th percentiles of ALMA-detected and undetected H-dropouts, respectively. Right panel, histogram showing the distribution of the 870-$\mu$m fluxes of ALMA-detected H-dropouts. {\bf b:} Main panel, the redshift and stellar masses are derived by template-fitting of their optical-to-NIR photometry, as described in Methods. The ALMA-undetected H-dropouts tend to be at slightly lower redshifts and have lower stellar masses, with a median redshift of $z_{\rm med}=3.8$ and stellar mass of $M_{*,\rm med}=10^{10.31} M_{\odot}$ while the ALMA-detected ones have $z_{\rm med} = 4.0$ and $M_{*,\rm {med}}=10^{10.56} M_{\odot}$. The error bars represent 1$\sigma$ uncertainties as determined from our SED-fitting procedure (Methods). Top and right panels, histogram of the redshift and stellar mass distributions of H-dropouts, respectively. The filled and open circles and their error bars denote the median redshift as well as the 16th and 84th percentiles of the ALMA-detected and undetected H-dropouts, respectively. 
\label{Fig:mass_pz}
}
\end{figure*}

\begin{figure*}
\centering
\includegraphics[width=0.5\textwidth]{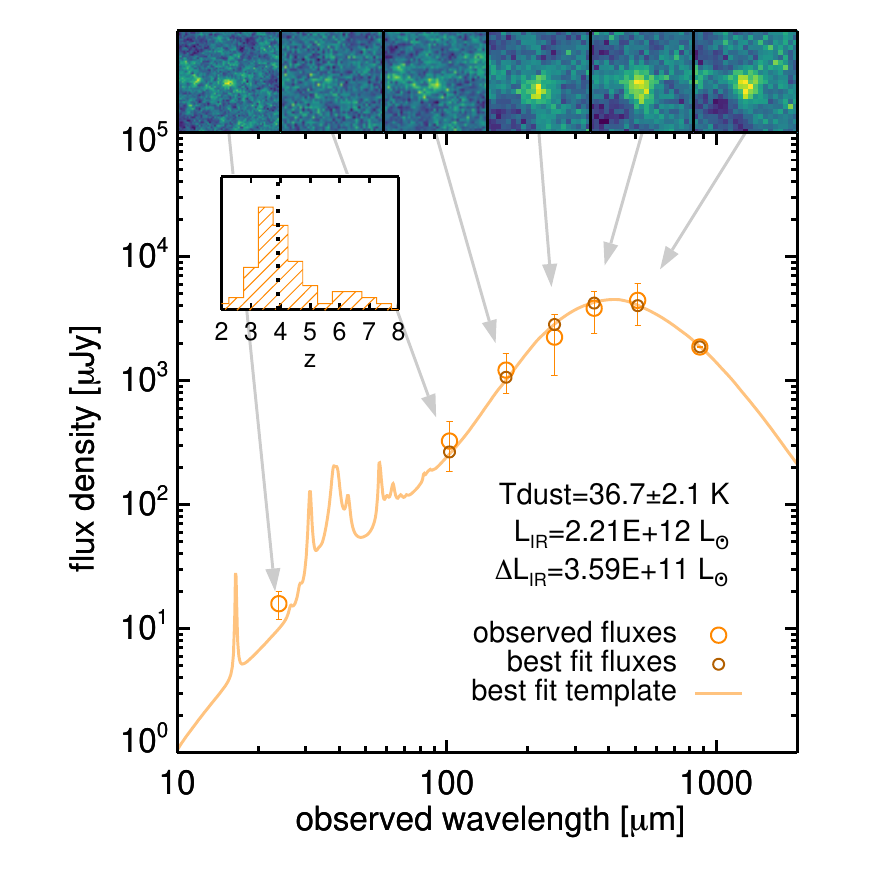}
\caption{\textbf{Stacked far-infrared SED of ALMA-detected H-dropouts.} The stacked IR SED is derived by median stacking of the 
Spitzer/24$\mu$m, Herschel/100$\mu$m, 160$\mu$m, 250$\mu$m, 350$\mu$m, 
500$\mu$m, and ALMA 870$\mu$m images of the 39 H-dropouts detected with 
ALMA. The measured fluxes from the stacked images and predicted fluxes from the best-fit model (solid line) are shown with the large and small open circles, respectively. Error bars (1$\sigma$) on the stacked SED are obtained from either bootstrapping or 
from the statistics of the residual map (whichever is largest, as 
described and validated elsewhere\cite{Schreiber:2015}). For the ALMA 
photometry, the error bar is the formal error on the mean ALMA flux, and 
is smaller than the data point on this figure. The stacked images are shown in the row of insets at the top, which are linked to their corresponding stacked photometric points by grey arrows. The inset histogram shows 
the photometric redshift distribution of the H-dropouts based on 
optical-to-NIR SED fitting, which shows a median redshift of $z \approx 4$. 
The infrared luminosity $L_{\rm IR}$ and dust temperature $T_{\rm dust}$ are derived from the 
best-fit SED at $z = 4$, the average redshift of the sample, using an 
empirical IR SED library calibrated on galaxies at $0<z<4$ 
(ref. \cite{Schreiber:2018a}). The uncertainty on the infrared luminosity ($\Delta L_{\rm IR}$)
accounts for uncertainty on the photometry and on the dust temperature, 
but not on the mean redshift of the sample.\label{Fig:stacked_IR}
}
\end{figure*}

\begin{figure*}
\centering
\includegraphics[width=0.95\textwidth]{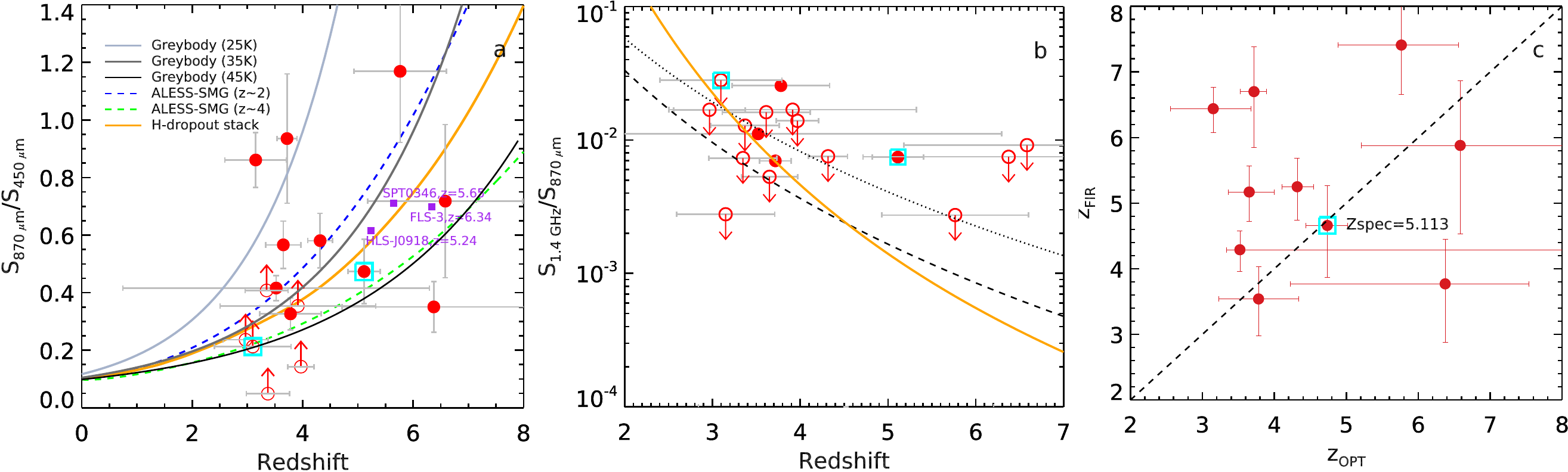}

\caption{\textbf{Photometric redshifts of H-dropouts.} {\bf a}, {\bf b}, $S_{870~\mu m}/S_{450~\mu m}$ ({\bf a}) and $S_{\rm 1.4~GHz}/S_{870~\mu m}$ ({\bf b}) colors versus redshifts for ALMA-detected H-dropouts in CANDELS-COSMOS; {\bf c}, comparison between redshifts derived from optical-to-NIR SEDs and from $S_{870~\mu m}/S_{450~\mu m}$ colors. {\bf a}, The redshifts are photometric redshifts derived from optical-to-NIR SED fitting except for the two sources denoted in cyan squares, which are spectroscopic redshifts derived from X-SHOOTER spectra. The $S_{870 \mu m}/S_{450 \mu m}$ color for galaxies undetected at 450 $\mu$m (S/N $<$ 2, open circles) are shown with their lower limits (using the $4\sigma$ upper limits at 450 $\mu$m). One of the spectroscopically confirmed galaxy with $z_{spec}$=3.097 is only marginally detected with $S_{870 \mu m} = 0.4\pm0.1$ mJy, below our conservative detection limit, but we also include it here for illustration. The  lines (see key) denote expected color evolution of different SED templates as a function of redshifts, including the stacked IR SED of the H-dropouts. We note that the $S_{870~\mu m}/S_{450~\mu m}$ color for both spectroscopically-confirmed sources are consistent with the average SED of ALESS $z=4$ SMGs. A few previously spectroscopically confirmed bright SMGs at $z > 5$ are shown by purple squares\cite{Riechers:2013,Combes:2012,Vieira:2013}.  {\bf b:} The 1.4 GHz flux is derived from 3 GHz assuming a spectral index of $\alpha=-0.8$. A 3 $\sigma$ upper limit of 7 $\mu$Jy is assigned to non-detections at 3 GHz, which are shown with open circles. The dotted and dashed  lines denote the relation between $S_{\rm 1.4~GHz}/S_{870~\mu m}$ and redshifts  for IR SEDs with spectral index in the submillimeter region of 3 (M82-like) and 3.5 (Arp220-like), respectively, as shown in ref. Carilli \& Yun (1999) \cite{Carilli:1999}. The same relation for the stacked IR SED of H-dropouts is also shown (orange line). {\bf c:} Comparison between submillimeter redshifts ($z_{\rm FIR}$), derived on the basis of their $S_{870~\mu m}/S_{450~\mu m}$ color and  their stacked IR SED (orange line in the left panel), and redshifts derived from optical-to-NIR SED fitting ($z_{\rm opt}$) for sources detected at both 450 $\mu$m and 870 $\mu$m. The cyan square denotes the source that is spectroscopically confirmed. Despite their large dispersion, both methods suggest that most of the $H-$dropouts are indeed at $z > 3$.
\label{Fig:submm_pz}
}
\end{figure*}

\begin{figure*}[!htb]
\centering
\includegraphics[width=0.95\textwidth]{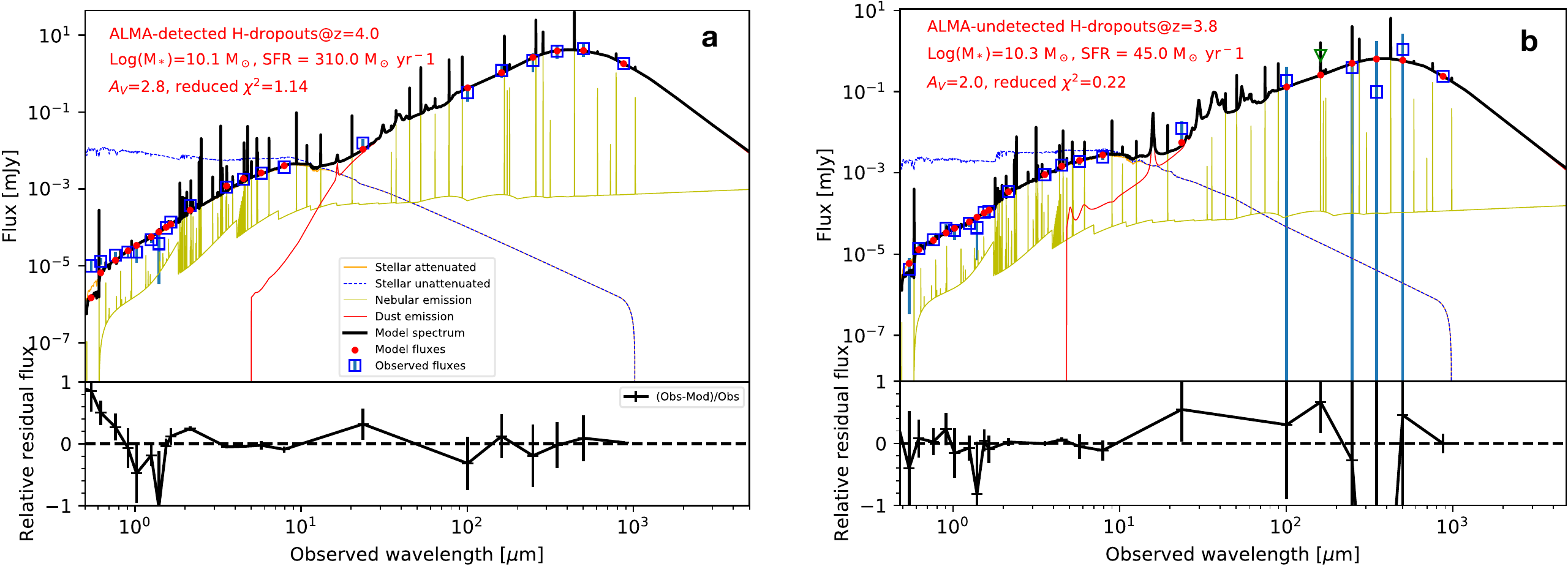}
\caption{\textbf{Full best-fit model of the stacked SEDs of ALMA-detected and -undetected H-dropouts.} {\bf a}, ALMA-detected; {\bf b}, ALMA-undetected. Here we show the best-fit SED templates obtained with the SED-fitting tool Cigale\cite{Boquien:2019}.  
We have adopted the BC03\cite{Bruzual:2003} library of single stellar populations and delayed star formation history model, with Draine \& Li ~\cite{Draine:2007a} models for the dust emission.
Nebular emission  based on CLOUDY templates was also included\cite{Inoue:2011}.
ALMA-undetected H-dropouts have much lower specific SFR (sSFR) compared to that of ALMA-detected ones. Error bars show standard measurement error (1$\sigma$.
\label{Fig:stacked_full}
}
\end{figure*}

\begin{figure*}
\centering
\includegraphics[width=0.8\textwidth]{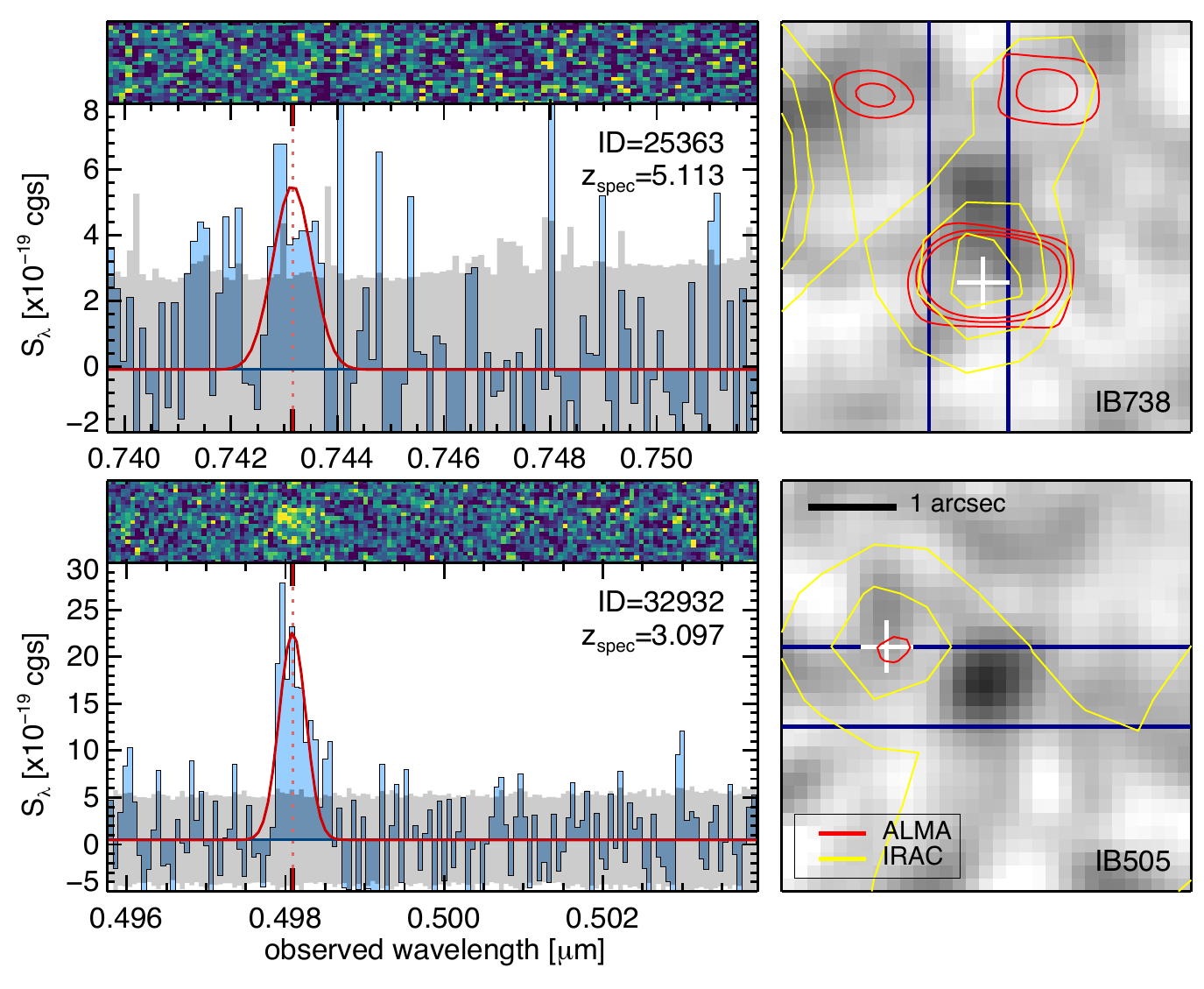}
\caption{ {\bf X-SHOOTER spectra of two spectroscopically-confirmed H-dropouts}. The two galaxies (with IDs 25363 and 32932) are shown on separate rows. Left, main panel, the observed spectra are shown on the left as black solid lines and blue shading, with uncertainties shown in the background as a gray shaded area. The best emission line model for Ly$\alpha$ is shown in red, and the centroid of the line is indicated with a vertical dotted line. The 2D spectrum is shown on the top, aligned with the 1D spectrum. Right, smoothed cutouts of the galaxies as observed on the Subaru medium band (IB738) where Ly$\alpha$ was detected. The X-SHOOTER slit is shown in blue, {\it Spitzer}--IRAC contours are shown in yellow, and ALMA contours are shown in red. The second galaxy with ID=32932 is only marginally detected with  $S_{870 \mu m}$ = $0.4\pm0.1$ mJy. The centroid of each dropout (determined from the IRAC image) is shown as a white cross.
\label{Fig:X-shooter}
}
\end{figure*}

\begin{figure*}
\centering
\includegraphics[width=0.95\textwidth]{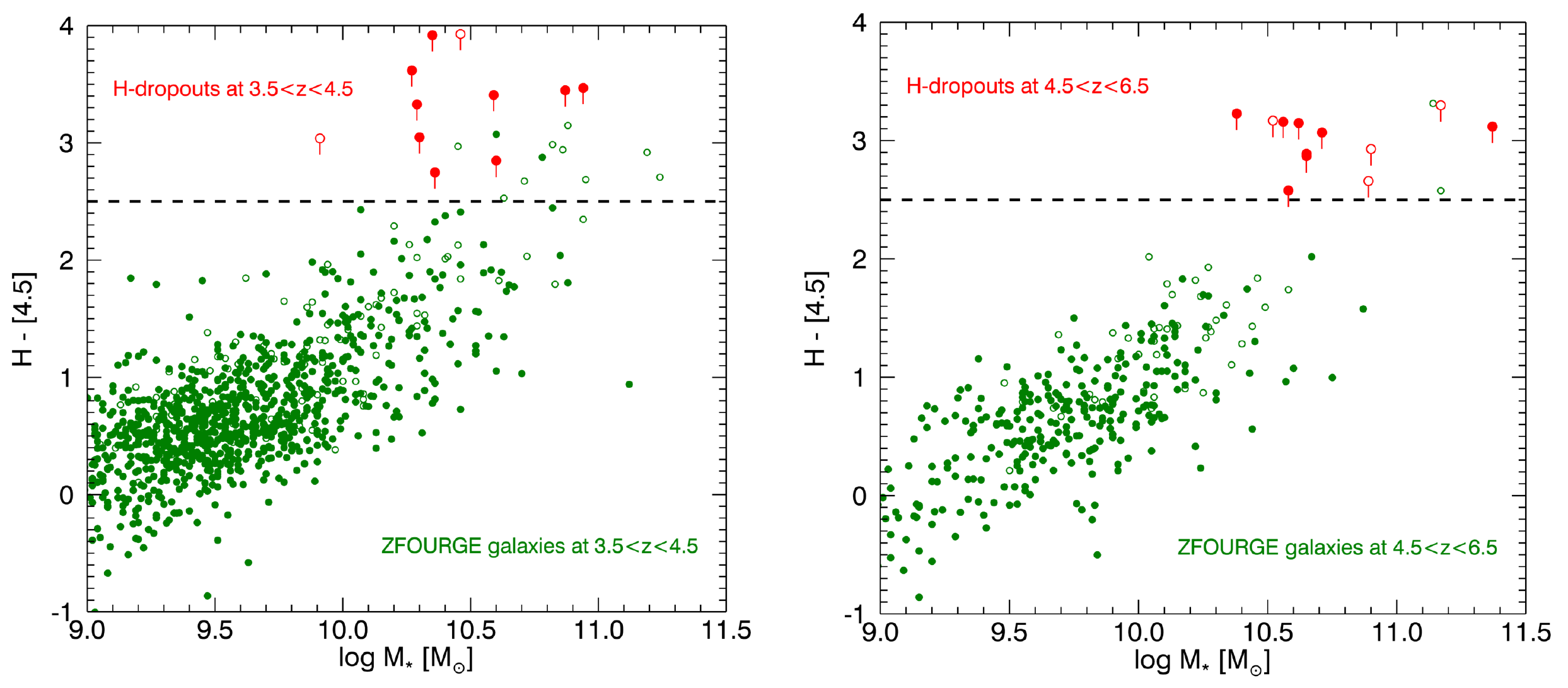}
\caption{ {\bf $H - [4.5]$ color versus stellar mass for massive galaxies at $3.5 < z < 6.5$}. Galaxies selected from the ZFOURGE catalog (left, $3.5 < z < 4.5$; right, $4.5 < z < 6.5$) with HST/F160W detections ($H < 27$) are shown in green while the H-dropouts selected in the same fields are shown in red. The $H - [4.5]$ color of the H-dropouts are shown by their lower limit assuming $H > 26.5 (5\sigma$). Quiescent and star-forming galaxies are shown by open and filled circles, respectively. Quiescent $H-$dropouts are defined as those undetected with ALMA while quiescent ZFOURGE galaxies are defined by their specific sSFR (based on SED fitting) with sSFR $< $0.3 Gyr$^{-1}$ and no MIPS 24 $\mu$m detections\cite{Spitler:2014}. 
\label{Fig:Xch2_lmass}
}
\end{figure*}

\begin{figure*}[!tbh]
\centering
\includegraphics[scale=0.5]{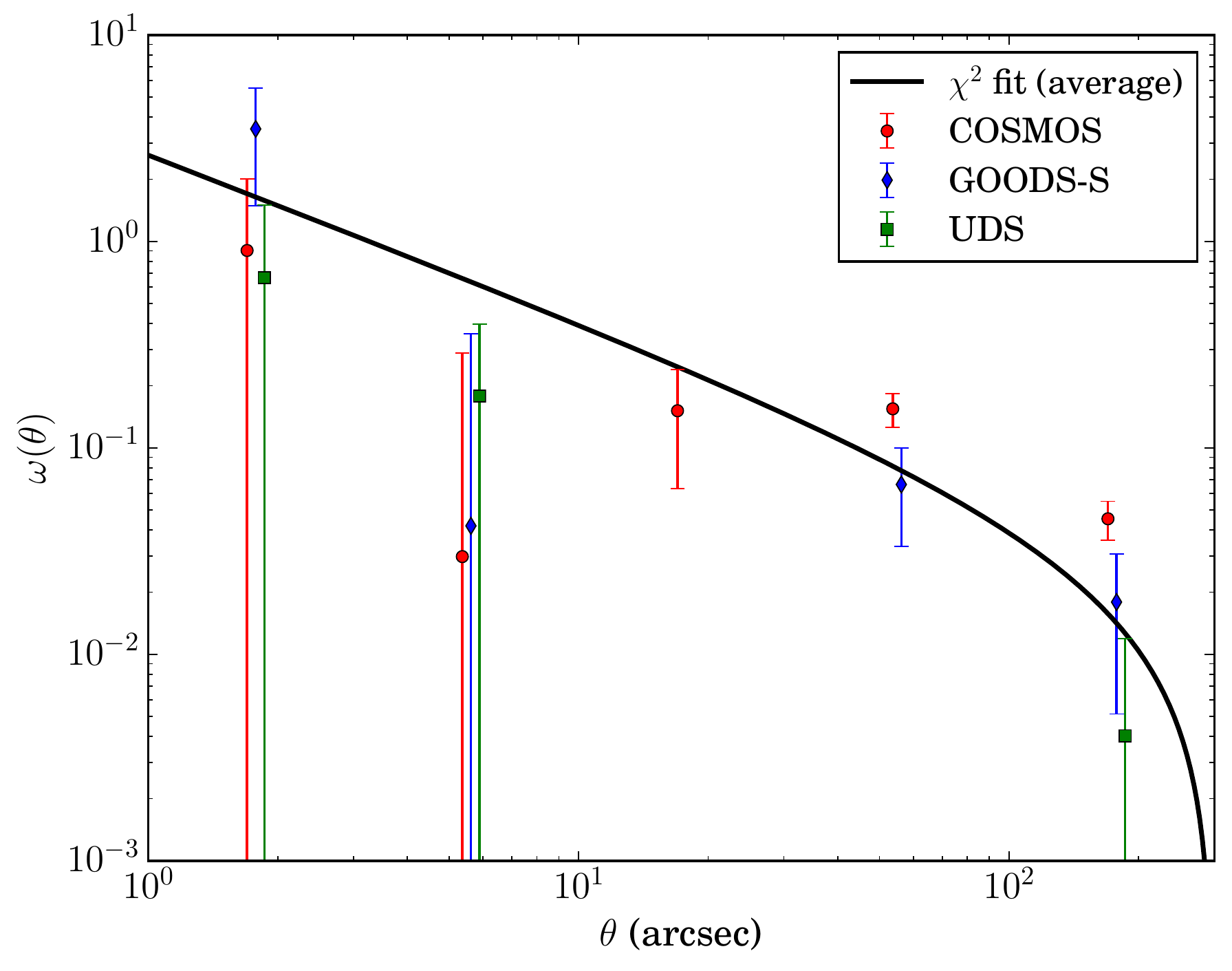}
\caption{\textbf{Angular cross-correlation function between H-dropouts and UV-selected galaxies at $3.5 < z < 5.5$}. The two-point angular cross-correlation function shown here, $\omega (\theta)$, is computed for the 39 ALMA-detected H-dropouts and  $\sim$6000 UV-detected (H-band) galaxies distributed in the same fields (CANDELS fields COSMOS, GOODS-S and UDS, see key). The solid black line is the best-fit line for the cross-correlation from the two-halo term  ($> 10\arcsec$ scale). The error bars are estimated from Poisson statistics. See Methods for details.
\label{Fig:cross_correlation}
}
\end{figure*}
\end{extendeddata}

\clearpage
\begin{addendum}
\item
This paper makes use of the following ALMA data: ADS/JAO.ALMA\#2015.1.01495.S, and ADS/JAO.ALMA\#2013.1.01292.S. ALMA is a partnership of ESO (representing its member states), NSF (USA), and NINS (Japan), together with NRC (Canada), NSC, ASIAA (Taiwan), and KASI (Republic of Korea), in cooperation with the Republic of Chile. The Joint ALMA Observatory is operated by ESO, AUI/NRAO, and NAOJ. This paper makes use of JCMT data from programs M16AL006,
MJLSC91, M11BH11A, M12AH11A, and M12BH21A. T. W. acknowledges the support by the NAOJ ALMA Scientific Research Grant Number 2017-06B, JSPS Grant-in-Aid for Scientific Research (S) JP17H06130, and funding from the European Union Seventh Framework Programme (FP7/2007-2013) under grant agreement No. 312725 (ASTRODEEP). X.S. acknowledges the support from NSFC 11573001, and National Basic Research Program 2015CB857005. C.-F.L. and W.-H.W. were supported by Ministry of Science and Technology of Taiwan Grant 105-2112-M-001-029-MY3.\\
\item[Author Contributions] 
T.W., C.S., and D.E. conceived the work, led the analysis and interpretation. T.W. proposed and carried out ALMA observations, reduced the ALMA data, measured SCUBA-2 fluxes and performed full SED fitting, and authored the majority of the text. C.S. conducted multiwavelength photometry and SED fitting, and carried out VLT/X-shooter observations and data reduction. Y.Yoshimura performed the clustering analysis. C.-F.L., W.-W.W. and X.S. contributed to the 450~$\mu$m photometry. K.K., Y. Yamaguchi, M.F., M.P., J.H., contributed to the overall interpretation of the results and various aspects of the analysis.

\item[Code availability]
The codes used to reduce ALMA and X-shooter data are public available. The codes used to model the optical-to-infrared SEDs, and to stack the optical and infrared images are accessible through github (https://github.com/cschreib). 

\item[Data availability]
Source data for the ALMA 870$\mu$m imaging are available through the ALMA archive. Optical-to-infrared imaging for all the galaxies in the sample are also public available through HST and Spitzer data archive. The other data that support the plots within this paper and other findings of this study are available from the corresponding author upon reasonable request.

 \item[Author Information] 
The authors declare that they have no
competing financial interests. Correspondence and requests for materials
should be addressed to T.W.~(email: twang.nju@gmail.com, ).

\end{addendum}

\end{document}